\documentclass[twocolumn,aps,prl,groupedaddress]{revtex4}

\usepackage{graphicx}
\usepackage{dcolumn}
\usepackage{bm}
\usepackage{amsmath,amssymb,amsfonts}

\begin{document}

\title{Interlayer Pairing Symmetry of Composite Fermions in Quantum Hall Bilayers}
\author{Hiroki Isobe}
\author{Liang Fu}
\affiliation{Department of Physics, Massachusetts Institute of Technology,
Cambridge, Massachusetts 02139, USA}

\begin{abstract}
We study the pairing symmetry of the interlayer paired state of composite fermions in quantum Hall bilayers. Based on the Halperin-Lee-Read (HLR) theory, the effect of the long-range Coulomb interaction and the internal Chern-Simons gauge fluctuation is analyzed with the random-phase approximation beyond the leading order contribution in small momentum expansion, and we observe that the interlayer paired states with a relative angular momentum $l=+1$ are energetically favored for filling $\nu=\frac{1}{2}+\frac{1}{2}$ and $\frac{1}{4}+\frac{1}{4}$. The degeneracy between states with $\pm l$ is lifted by the interlayer density-current interaction arising from the interplay of the long-range Coulomb interaction and the Chern-Simons term in the HLR theory. 
\end{abstract}


\maketitle

Quantum Hall systems with even-denominator filling fractions are well described by composite fermions (CFs) \cite{Jain}. A CF  in two dimensions is composed of an electron with an even number of magnetic fluxes attached via the Chern-Simons gauge field. 
The attached fluxes cancel the external magnetic field on average, thus leading to a well-defined Fermi surface of CFs as theorized by Halperin, Lee, and Read \cite{HLR}. 

In quantum Hall bilayer systems, quantized Hall conductances, indicative of incompressible states, are observed when each layer is at even-denominator filling fractions and two layers are separated by a short distance. Such systems are realized in a single wide quantum well \cite{wide}, double quantum wells \cite{double}, and more recently, bilayer graphene \cite{graphene1,graphene2,graphene3,graphene4}. 
Tunneling spectroscopy \cite{tunnel1,tunnel2}, Hall drag \cite{hall}, and counterflow measurements \cite{counter1,counter2} demonstrate the formation of an exciton superfluid phase for small layer distances \cite{exciton1,exciton2,exp_review}. 
On the other hand, the bilayer system is described by two composite Fermi liquids with interlayer interactions at large distance. 
From a theoretical viewpoint, Bonesteel \textit{et al.} \cite{Bonesteel1,Bonesteel2} showed that such a system is unstable to Cooper pairing between CFs on the two different layers. 
The pairing interaction arises from the long-range Coulomb interaction and fluctuations of the Chern-Simons gauge fields. 
Using the random-phase approximation (RPA) for the gauge field propagator, Refs.~\cite{Bonesteel1,Bonesteel2} derived the most singular part of the pairing interaction. 
As recognized by the authors, at this level of approximation, pairing interactions in all angular momentum channels are degenerate.

In this Letter, we study the energetically favored pairing symmetry of bilayer quantum Hall systems due to the effective interaction between CFs obtained by the RPA. 
We go beyond the previous analyses to include the effect of the time-reversal breaking external magnetic field on the effective interaction between CFs. 
This effect appears through an interlayer density-current interaction mediated by the Chern-Simons gauge field. 
The resulting pairing interaction between CFs lifts the degeneracy between pairings in angular momentum $+l$ and $-l$ channels.
We show that the interlayer paired state with a relative angular momentum $l=+1$ is favored at filling $\nu=\frac{1}{2}+\frac{1}{2}$ and $\frac{1}{4}+\frac{1}{4}$. 
Here, we define the angular momentum of the Moore-Read Pfaffian state \cite{MR} as $l=+1$.

\begin{figure}
\centering
\includegraphics[width=\hsize]{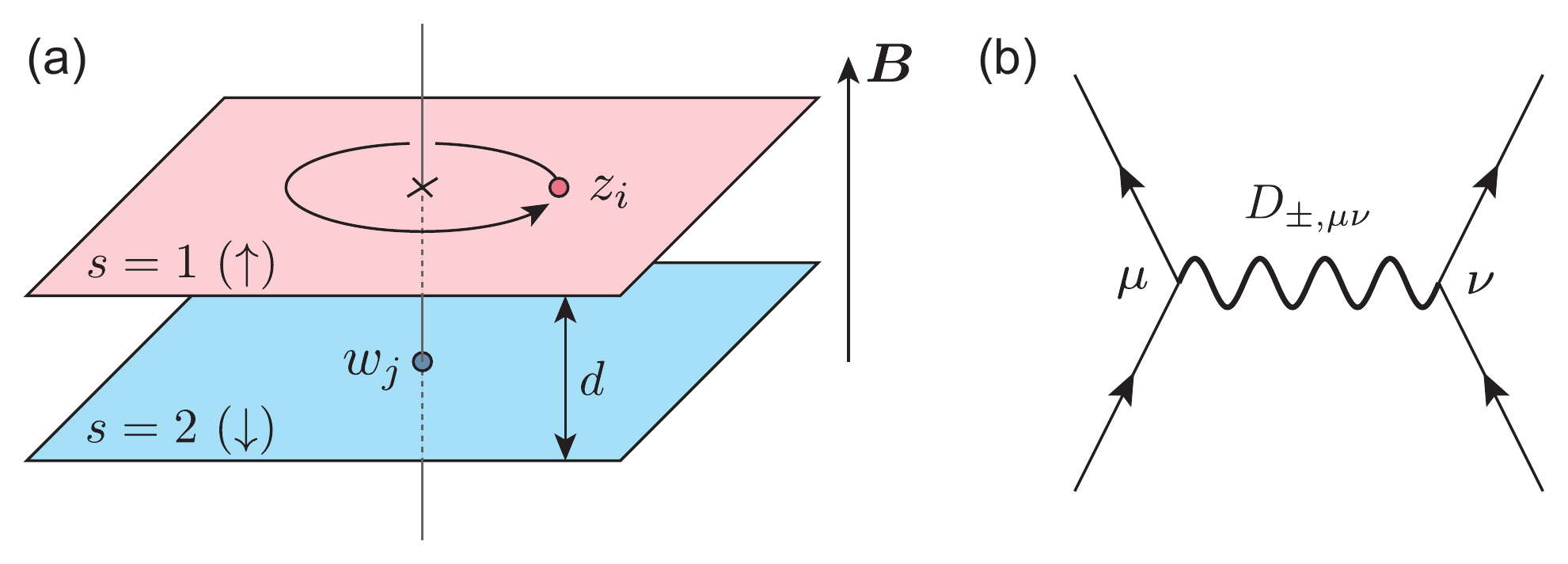}
\caption{
(a) Geometry of the bilayer system. The magnetic field $\bm{B}$ is applied upward through the two layers with the distance $d$. 
An interlayer paired state with a relative angular momentum $l$ gives a winding phase $2\pi l$ when one moves a CF counterclockwise around another in the other layer.
(b) Effective interaction for CFs. $\mu=0$ $(1)$ at a vertex means a coupling between the density (current) fluctuation of CFs and the Chern-Simons gauge field. 
}
\label{fig:int}
\end{figure}

\textit{Model.}---%
We consider a bilayer system of CFs with layer spacing $d$ in the presence of the long-range Coulomb interaction [Fig.~\ref{fig:int}(a)].  We assume that the filling fraction is the same for both layers. In the imaginary time formalism, the partition function is 
$Z=\int \prod_s D\psi_s^\dagger D\psi_s D\bm{a}^{(s)} Da_0^{(s)} e^{-S}$, 
with the action $S=\int_0^\beta d\tau \int d^2 r \mathcal{L} (\bm{r},\tau)$. 
The Lagrangian density $\mathcal{L}$ is given by \cite{Bonesteel1,Bonesteel2,Kim}
\begin{align}
& \mathcal{L}(\bm{r},\tau) \notag \\
=& \sum_{s} \bigg\{ \psi_s^{\dagger}(\bm{r},\tau) \left[\partial_{\tau}+ia_{0}^{(s)}(\bm{r},\tau) \right] \psi_s(\bm{r},\tau) \notag \\
& +\frac{1}{2m^{*}}\psi_s^{\dagger}(\bm{r},\tau) \left[ -i\nabla-\bm{a}^{(s)}(\bm{r},\tau) +e\bm{A}(\bm{r}) \right]^{2}\psi_s(\bm{r},\tau) \notag \\
& -\mu \psi_s^{\dagger}(\bm{r},\tau) \psi_s(\bm{r},\tau)
\bigg\} \notag \\
& -\sum_{ss'}\frac{i}{2\pi}K_{ss'}^{-1}a_{0}^{(s)}(\bm{r},\tau) \hat{z}\cdot [\nabla\times \bm{a}^{(s')}(\bm{r},\tau)] \notag \\
& +\frac{1}{2}\sum_{ss'}\int d^{2}r'\delta\rho_s(\bm{r},\tau)V_{ss'}(\bm{r}-\bm{r}')\delta\rho_{s'}(\bm{r}',\tau), 
\end{align}
where $\psi_s$ represents the CF field with $s=1,2$ (or $\uparrow, \downarrow$) being a layer index,  $m^*$ is the effective mass of the CFs, $\bm{a}^{(s)}$ and $a_0^{(s)}$ are the Chern-Simons gauge fields, and $\bm{A}$ is the $U(1)$ gauge field for the uniform external magnetic field $B$ along the $z$ direction. Here, we employ units where $\hbar=c=1$, and the Coulomb gauge for the Chern-Simons gauge field; $\nabla\cdot\bm{a}^{(s)}=0$.  The electron charge is $-e$. 
The filling fraction of each layer is $2\pi n_e/(eB)$, where $n_e$ is the electron density, and $\mu$ is the chemical potential. 
The energy dispersion is $\epsilon_{\bm{k}} = k^2/(2m^*)$, and the Fermi wave vector $k_F$ is given by $k_F=\sqrt{4\pi n_e}=\sqrt{2\nu}/l_0$, where the magnetic length is $l_0=(eB)^{-1/2}$. 
The Coulomb interaction $V_{ss'}(\bm{r})=e^2/(\varepsilon r)$ $(s=s')$ or $e^2/(\varepsilon \sqrt{r^2+d^2})$ $(s\neq s')$ \cite{screen} acts on the density fluctuation $\delta\rho_{s}(\bm{r},\tau)=\psi_{s}^{\dagger}(\bm{r},\tau)\psi_{s}(\bm{r},\tau)-n_e$. 
The elements of the $K$ matrix are taken as $K_{11}=K_{22}=\tilde{\phi}$ and $K_{12}=K_{21}=0$ with the integer $\tilde{\phi}$ corresponding to the number of fluxes attached to an electron. This is confirmed by integrating out $a_0^{(s)}$, to obtain the constraint $\psi_s^\dagger \psi_s = \hat{z}\cdot\nabla\times\bm{a}^{(s)}/(2\pi\tilde{\phi})$. 
Note that the sign of $\tilde{\phi}$ represents the direction of the magnetic field, and it changes by time-reversal operation; we take $\tilde{\phi}>0$ in the following analysis to make the direction of the magnetic field point upward.  
The filling fraction of each layer is $\tilde{\phi}^{-1}$, so that the CFs feel effectively no magnetic field on average.  The density fluctuation is given by
\begin{equation}
\delta\rho_s (\bm{r},\tau) =\frac{1}{2\pi\tilde{\phi}} \hat{z}\cdot\nabla\times [\bm{a}^{(s)}(\bm{r},\tau) -e\bm{A}(\bm{r})]. 
\end{equation}

\textit{Effective interaction.}---%
The effective action for the gauge field is obtained by a saddle-point approximation with expansion about the point where $a_0^{(s)}=0$ and $\bm{a}^{(s)}-e\bm{A}=0$. 
With the Coulomb gauge condition, the gauge fluctuation in the spatial part can be written by $a_1^{(s)} (\bm{q},i\omega_m) = \hat{z}\cdot \left\{\hat{\bm{q}}\times \left[ \bm{a}^{(s)}(\bm{q}, i\omega_m)-e\bm{A}(\bm{q})\right]\right\}$, where $\omega_m=2m \pi T$ is a bosonic Matsubara frequency. Up to the second order in the gauge field, the effective action is
\begin{align}
S_\text{eff} =& \frac{1}{2}T\sum_{\omega_m}\int\frac{d^2q}{(2\pi)^2} \sum_{ss'} \sum_{\mu,\nu=0,1} a_\mu^{(s)} (\bm{q},i\omega_m) \notag \\
& \times D^{-1}_{s\mu,s'\nu} (q,i\omega_m) a_\nu^{(s')}(-\bm{q},-i\omega_m). 
\end{align}
It is useful for later analysis to decompose the gauge field into in-phase and out-of-phase fluctuations $a_\mu^{(\pm)}=(a_\mu^{(1)}\pm a_\mu^{(2)})/\sqrt{2}$, with the corresponding propagator $D_{\pm,\mu\nu}$. 
$D^{-1}_{\pm,\mu\nu}$ is obtained with the RPA \cite{Bonesteel1,Bonesteel2,Bonesteel3,supplement}, whose singular terms for $\omega/\epsilon_{F}\ll (q/k_{F})^2 \ll1$ and $q\ll d^{-1}$ are 
\begin{subequations}
\label{eq:small-q}
\begin{gather}
D_{-,11}(q,i\omega_m)\approx-\frac{1}{\tilde{\chi}_{d}q^{2}+\frac{k_{F}}{2\pi}\frac{|\omega_m|}{q}},\\
D_{+,11}(q,i\omega_m)\approx-\frac{1}{\frac{e^{2}}{\pi\varepsilon\tilde{\phi}^{2}}q+\frac{k_{F}}{2\pi}\frac{|\omega_m|}{q}},\\
D_{-,01}(q,i\omega_m)=D_{-,10}(q,i\omega_m)\approx\frac{1}{\tilde{\chi}_{d}q^{2}+\frac{k_{F}}{2\pi}\frac{|\omega_m|}{q}}\frac{q}{m^{*}\tilde{\phi}},
\end{gather}
\end{subequations}
with $\tilde{\chi}_d=\frac{1}{24\pi m^*}+\frac{e^2 d}{2\pi\varepsilon\tilde{\phi}^2}+\frac{1}{2\pi m^* \tilde{\phi}^2}$.

From the effective action and the gauge propagator, the effective interaction between the CFs [Fig.~\ref{fig:int}(b)] is obtained by
\begin{align}
\mathcal{V}=&\frac{1}{2}\sum_{s_1 s_2 s_3 s_4}\psi_{s_1}^{\dagger}(\bm{k}+\bm{q},i\epsilon_n+i\omega_m) \psi_{s_2}^\dagger (\bm{k}'-\bm{q},i\epsilon'_n-i\omega_m) \notag \\
&\times V_{s_1 s_2 s_3 s_4}^{\text{eff}}(\bm{k},\bm{k}',\bm{q},i\omega_m) \psi_{s_3}(\bm{k}',i\epsilon'_n) \psi_{s_4} (\bm{k},i\epsilon_n),
\end{align}
where $\epsilon_n=(2n+1)\pi T$ is a fermionic Matsubara frequency, and the matrix element is 
\begin{align}
&V_{s_{1}s_{2}s_{3}s_{4}}^{\text{eff}}(\bm{k},\bm{k}',\bm{q},i\omega_{m}) \notag \\
=&-\sum_{\mu,\nu=0,1}M_{\mu\nu}(\bm{k},\bm{k}',\hat{\bm{q}}) \bigl[D_{+,\mu\nu}(q,i\omega_{m})(\sigma_{0})_{s_{1}s_{4}}(\sigma_{0})_{s_{2}s_{3}} \notag \\
&+D_{-,\mu\nu}(q,i\omega_{m})(\sigma_{3})_{s_{1}s_{4}}(\sigma_{3})_{s_{2}s_{3}} \bigr],
\label{eq:effective}
\end{align}
with 
\begin{equation}
M_{\mu\nu}(\bm{k},\bm{k}',\hat{\bm{q}})=\frac{1}{2}\begin{pmatrix}1 & -i\frac{\hat{z}\cdot(\hat{\bm{q}}\times\bm{k}')}{m^*}\\
i\frac{\hat{z}\cdot(\hat{\bm{q}}\times\bm{k})}{m^*} & \frac{(\hat{\bm{q}}\times\bm{k})\cdot(\hat{\bm{q}}\times\bm{k}')}{m^{*2}}
\end{pmatrix}_{\mu\nu},
\end{equation}
which dictates the coupling of the Chern-Simons gauge field fluctuation to the CFs. 
Here, the Pauli matrix $\sigma_\alpha$ $(\alpha=0,...,3)$ acts on layer indices. 

The dominant contribution in the effective interaction at small $q$ comes from the out-of-phase fluctuation of the current-current correlation $D_{-,11}$. Preceding analysis explained the existence of a stable interlayer paired state by taking only the current-current propagator $D_{\pm,11}$ \cite{Bonesteel1,Bonesteel2}. 
However, this is not enough to examine the stable pairing symmetry because time-reversal symmetry breaking by the external magnetic field is absent. To this end, it is necessary to include the density-current propagators $D_{\pm,01}$ and $D_{\pm,10}$, which are induced by the Chern-Simons term and change sign under time reversal ($\tilde{\phi}\to -\tilde{\phi}$). In the following analysis, we include all terms in the effective interaction \eqref{eq:effective} on an equal footing. 

\textit{Pairing symmetry and wave functions}.---%
We investigate the stable pairing state using the framework of the Eliashberg theory. Here, the Green's function of the CFs in the Nambu space is written as 
\begin{align}
G^{-1} (\bm{k},i\epsilon_n) =
\begin{pmatrix}
(i\epsilon_n Z_{n}-\xi_{\bm{k}}) \sigma_{0} & \hat{\phi}_n(\bm{k})\\
\hat{\phi}_n^\dagger (\bm{k}) & (i\epsilon_n Z_{n}-\xi_{\bm{k}}) \sigma_{0}
\end{pmatrix},
\end{align}
where $Z_n$ is the quasiparticle residue, $\hat{\phi}_n(\bm{k})$ is the anomalous self-energy, and $\xi_{\bm{k}}=\epsilon_{\bm{k}}-\mu$. The gap function is given by $\Delta_n(\bm{k}) =\hat{\phi}_n(\bm{k})/Z_n$. 
We focus on fully gapped interlayer paired states.  With the in-plane rotational symmetry, we have 
$\hat{\phi}^{(l)}_n(\bm{k}) = \phi_n (i\sigma_2) e^{il\theta_{\bm{k}}}$ (even $l$), or 
$\hat{\phi}^{(l)}_n(\bm{k}) = \phi_n (i\sigma_3 \sigma_2) e^{il\theta_{\bm{k}}}$ (odd $l$), 
where $l$ is the relative angular momentum and $\theta_{\bm{k}}$ is the azimuth of $\bm{k}$ \cite{triplet}.

The Green's function $G(\bm{k},i\epsilon_n)$ yields the effective action for the CFs. Recalling the BCS theory, we obtain the ground state of the CFs as 
\begin{equation}
|\Psi_\text{CF}\rangle \propto \prod_{\bm{k}} (1+g_{\bm{k}} c^\dagger_{\bm{k}\uparrow} c^\dagger_{-\bm{k}\downarrow}) |0\rangle. 
\end{equation}
$|0\rangle$ is the vacuum containing no particles, $c^\dagger_{\bm{k}s}$ creates a CF of momentum $\bm{k}$ on layer $s$, and the function $g_{\bm{k}}$ is $g_{\bm{k}}=\phi_n e^{il\theta_{\bm{k}}}/(\xi_{\bm{k}}+E_{\bm{k}})$ with $E_{\bm{k}}=\sqrt{\xi_{\bm{k}}^2+|\phi_n|^2}$ \cite{supplement}. 
The wave function of a system with $N$ electrons in each layer is obtained by 
\begin{equation}
\Psi_\text{CF} (\{\bm{r}_\uparrow\}, \{\bm{r}_\downarrow\}) = \det[ g(\bm{r}_{i\uparrow}, \bm{r}_{j\downarrow})],
\end{equation} 
where $g(\bm{r}_{i\uparrow}, \bm{r}_{j\downarrow})$ is the Fourier transform of $g_{\bm{k}}$; $g(\bm{r}_{i\uparrow}, \bm{r}_{j\downarrow}) = L^{-2} \sum_{\bm{k}} g_{\bm{k}} e^{i\bm{k}\cdot(\bm{r}_{i\uparrow}-\bm{r}_{j\downarrow})}$ ($L^2=$ the area of the system). 

The electron wave function for an interlayer paired state generally has a form 
\begin{align}
\label{eq:wave}
\Psi(\{z\}, \{w\}) =& \mathcal{P}_\text{LLL} \prod_{i<j} (z_i-z_j)^{\tilde{\phi}} \prod_{i'<j'} (w_{i'}-w_{j'})^{\tilde{\phi}} \notag \\
&\times \det[ g(z_i, w_j) ], 
\end{align}
where $\mathcal{P}_\text{LLL}$ is the projection operator onto the lowest Landau level. 
Here, we introduce the complex representations of the coordinate $z_i=x_{i\uparrow}-iy_{i\uparrow}$ and $w_j=x_{j\downarrow}-iy_{j\downarrow}$ \cite{definition}. 
The first two terms in the right-hand side describe the fluxes attached to the electrons. With an even $\tilde{\phi}$, this bosonic part corresponds to the Halperin $(\tilde{\phi},\tilde{\phi},0)$ state \cite{Halperin}. 
For an interlayer paired state with an angular momentum $l$, we have $g(z_i,w_j)\sim (z_i-w_j)^{-l}$ in short distances \cite{supplement}, which produces a winding phase $2\pi l$; see Fig.~\ref{fig:int}(a). 
Using the Cauchy identity, the paired CF part can be regarded as the $(l,l,-l)$ state for a weak-pairing case \cite{RG}.

\begin{figure*}
\centering
\includegraphics[width=0.9\hsize]{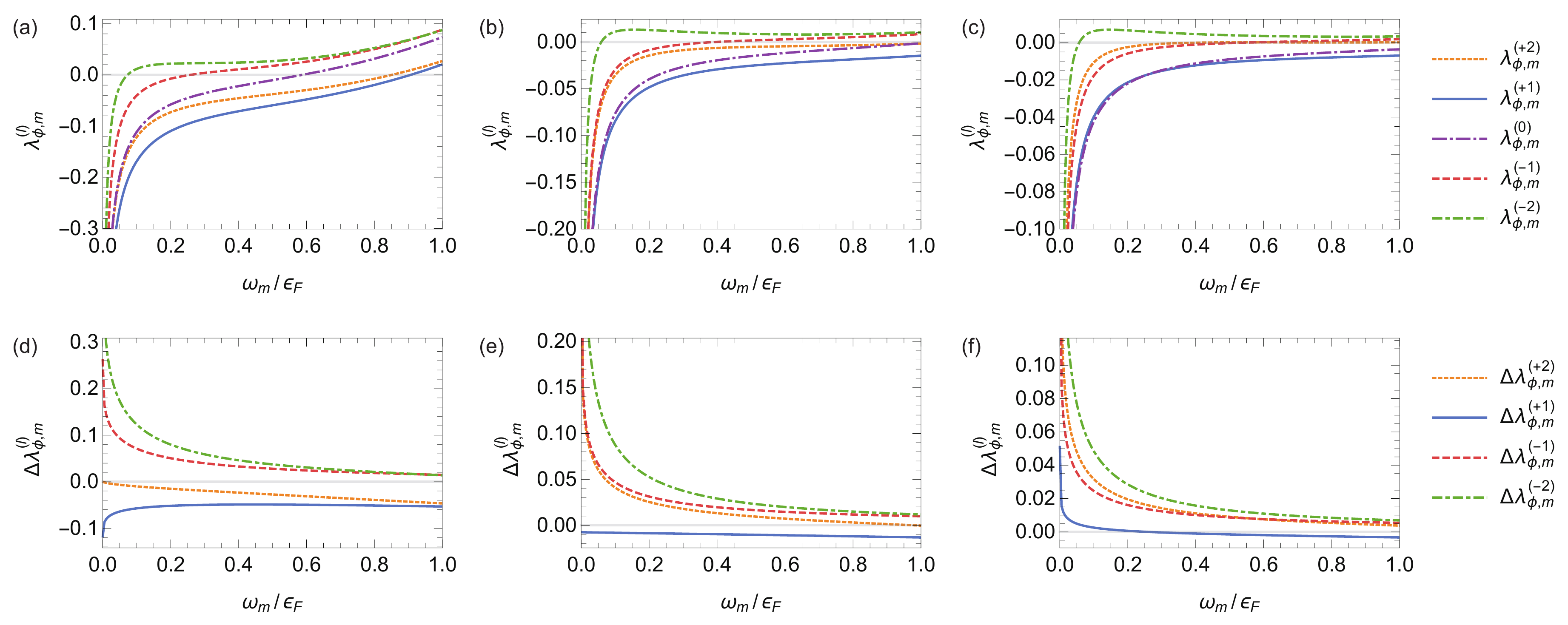}
\caption{
Frequency dependence of (a)--(c) the effective coupling constants $\lambda_{\phi,m}^{(l)}$ and (d)--(f) the difference $\Delta\lambda_{\phi,m}^{(l)}=\lambda_{\phi,m}^{(l)}-\lambda_{\phi,m}^{(0)}$. 
We set the filling fraction in (a) and (d) to $\nu=\frac{1}{2}+\frac{1}{2}$, in (b) and (e) to $\nu=\frac{1}{4}+\frac{1}{4}$, and in (c) and (f) to $\nu=\frac{1}{6}+\frac{1}{6}$. 
The ratio of the Coulomb energy to the Fermi energy is $(e^2/\varepsilon l_0)/\epsilon_F=1$ and the layer spacing is $k_F d=1$. At filling $\nu=\frac{1}{2}+\frac{1}{2}$ and $\frac{1}{4}+\frac{1}{4}$, the $l=+1$ state is favored for all frequencies. In contrast, the $l=0$ pairing is stable for low frequencies at $\nu =\frac{1}{6}+\frac{1}{6}$. 
}
\label{fig:numerical}
\end{figure*}

\textit{Energetics of paired states.}---%
The quasiparticle residue $Z_n$ receives a correction from the exchange interaction 
\begin{align}
&V_\text{ex} (\bm{k},\bm{q},i\omega_m) \notag \\
=& -\sum_{\mu\nu} M_{\mu\nu} (\bm{k},\bm{k}+\bm{q},\hat{\bm{q}}) \left[D_{+,\mu\nu}(q,i\omega_{m})+D_{-,\mu\nu}(q,i\omega_{m}) \right],
\end{align}
and the anomalous self-energy $\hat{\phi}_n (\bm{k})$ is related to the interaction in the Cooper channel
\begin{align}
&V_c (\bm{k},\bm{q},i\omega_m) \notag \\
=& \sum_{\mu\nu} M_{\mu\nu} (\bm{k},-\bm{k}-\bm{q},\hat{\bm{q}}) \left[D_{+,\mu\nu}(q,i\omega_{m})-D_{-,\mu\nu}(q,i\omega_{m}) \right]. 
\end{align}
In the Cooper channel, $D_+$ and $D_-$ have the different signs, which reflects the fact that the two layers have the opposite $a^{(-)}$ gauge charges. 
Importantly, off-diagonal terms in $M_{\mu\nu}$, which correspond to density-current interactions and break time-reversal symmetry, affect only $V_c$.

We assume $\Delta_n (\bm{k}) \ll \epsilon_F$, so that the pairing occurs only on the Fermi surface. 
Then we define the effective coupling constants for $Z_n$ and $\hat{\phi}_n^{(l)} (\bm{k})$ as $\lambda_{Z,m}$ and $\lambda_{\phi,m}^{(l)}$, respectively: 
\begin{gather}
\lambda_{Z,m} = \int \frac{d^2q}{(2\pi)^2} \delta(\xi_{\bm{k}+\bm{q}}) V_\text{ex} (\bm{k},\bm{q},i\omega_m),  \notag \\
\lambda_{\phi,m}^{(l)} = \int\frac{d^2q}{(2\pi)^2} \delta(\xi_{\bm{k}+\bm{q}}) V_c (\bm{k},\bm{q},i\omega_m) \left( 1+\frac{q}{k_F} e^{i\theta_{\bm{q}}} \right)^l,
\label{eq:eff_phi}
\end{gather}
with the condition $|\bm{k}|=k_F$.
The effective coupling constants are related to the Eliashberg equations \cite{supplement}
\begin{gather}
\left(1-Z_{n}\right)\epsilon_{n}=-\pi T\sum_{\omega_{m}}\frac{\lambda_{Z,m} Z_{n+m}(\epsilon_{n}+\omega_{m})}{\sqrt{Z_{n+m}^{2}(\epsilon_{n}+\omega_{m})^{2}+|\phi_{n+m}^{(l)}|^{2}}}, \notag \\
\phi_{n}^{(l)}=-\pi T\sum_{\omega_{m}}\frac{\lambda_{\phi,m}^{(l)} \phi_{n+m}^{(l)}}{\sqrt{Z_{n+m}^{2}(\epsilon_{n}+\omega_{m})^{2}+|\phi_{n+m}^{(l)}|^{2}}}.
\end{gather}

The stable pairing symmetry can be examined from $\lambda_{\phi,m}^{(l)}$, shown in Figs.~\ref{fig:numerical}(a)--\ref{fig:numerical}(c). 
The integrations in Eq.~\eqref{eq:eff_phi} have divergences as $q\to 0$, and a cutoff $q_c=10^{-5}k_F$ is introduced to cure them \cite{supplement}.  
Negative values of $\lambda_{\phi,m}^{(l)}$ mean attractive interaction at $\omega_m$, and the stable pairing symmetry will be the one that has the strongest attractive interaction.  

The differences of the effective coupling constants $\Delta\lambda_{\phi,m}^{(l)} = \lambda_{\phi,m}^{(l)} - \lambda_{\phi,m}^{(0)}$ clearly display the stable pairing symmetry [Figs.~\ref{fig:numerical}(d)--\ref{fig:numerical}(f)]. 
They do not have a singularity, and hence the cutoff is not necessary. 
We find that the $l=+1$ state is favored at all frequencies when the filling fraction is $\nu=\frac{1}{2}+\frac{1}{2}$ or $\frac{1}{4}+\frac{1}{4}$. The result suggests that a Cooper pair in the interlayer paired phase has an angular momentum $l=+1$. 
In contrast, the $l=0$ state is favored at small frequencies for $\nu=\frac{1}{6}+\frac{1}{6}$. 
We note that the degeneracy of the states with $\pm l$ is lifted since the time-reversal symmetry is broken due to the coupling of the density and current fluctuations via the Chern-Simons term.

\begin{figure}
\centering
\includegraphics[width=\hsize]{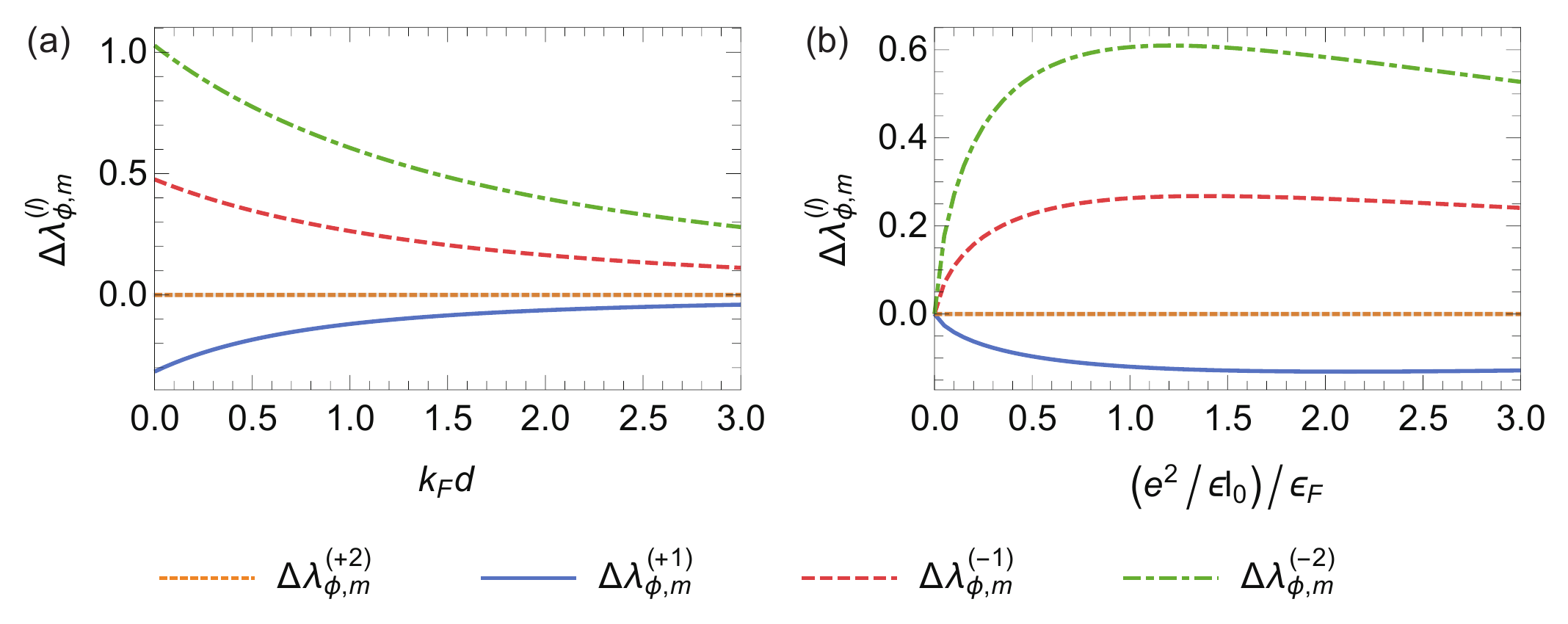}
\caption{
(a) Layer spacing dependence of $\Delta\lambda_{\phi,m}^{(l)}$.  We set $(e^2/\varepsilon l_0)/\epsilon_F=1$ and $\omega_m=0$ at $\nu=\frac{1}{2}+\frac{1}{2}$. Reducing the spacing makes the interaction strength stronger. 
(b) Effective mass dependence of $\Delta\lambda_{\phi,m}^{(l)}$. Note $m^*\propto (e^2/\varepsilon l_0)/\epsilon_F$. We set $k_F d=1$ and $\omega_m=0$ at $\nu=\frac{1}{2}+\frac{1}{2}$.
In both cases, the ordering of $\Delta\lambda_{\phi,m}^{(l)}$ does not change. At $\nu=\frac{1}{2}+\frac{1}{2}$, the $l=+1$ pairing is favored at any cases. 
$\Delta\lambda_{\phi,0}^{(+2)}$ identically vanishes for $\tilde{\phi}=2$. See also Eq.~\eqref{eq:expansion}. 
}
\label{fig:length}
\end{figure}

The layer spacing and the effective mass dependences of $\Delta\lambda_{\phi,m}^{(l)}$ at $\nu=\frac{1}{2}+\frac{1}{2}$ are also examined (Fig.~\ref{fig:length}). As the layer spacing $d$ decreases, the differences of $\Delta\lambda_{\phi,m}^{(l)}$ increase, but the ordering remains unchanged. Controlling $(e^2/\varepsilon l_0)/\epsilon_F$, proportional to the effective mass $m^*$, also does not change the ordering of $\Delta\lambda_{\phi,m}^{(l)}$.  Similar results for other filing fractions are provided in the Supplemental Material \cite{supplement}. 

\textit{Discussions.}---%
It is instructive to examine $\lambda_{\phi,m}^{(l)}$ using the small-$q$ expansion of $V_c (\bm{k},\bm{q},i\omega_m)$. 
A formation of a paired state is explained by considering the singular terms at $\omega_m =0$: 
\begin{align}
\lambda_{\phi,0}^{(l)}=\frac{1}{(2\pi)^2}\frac{k_F}{m^*} \int_0^{2k_F} dq \left( -\frac{1}{\tilde{\chi}_d q^2} +\frac{1}{\frac{e^2}{\pi\varepsilon\tilde{\phi}^2}q} +O(q^0) \right), 
\label{eq:expansion_sing}
\end{align}
which is independent of pairing symmetries. 
These singularities are smeared at finite frequencies, see Eq.~\eqref{eq:small-q}. 
$\lambda_{Z,m}$ also has the similar structure, but it does not disturb a formation of pairing \cite{wang}. 
The first term represents attractive interaction originated from the out-of-phase fluctuation $a_1^{(-)}$  because $a_\mu^{(-)}$ sees the CFs in the different layers as oppositely charged. The second term comes from the in-phase fluctuation $a_1^{(+)}$, which gives repulsive interaction. 

In Eq.~\eqref{eq:expansion_sing}, the effect of the Chern-Simons term and hence time-reversal symmetry breaking is absent in the singular terms. The difference is found from $q^0$ order; we obtain
\begin{align}
\Delta\lambda_{\phi,0}^{(l)} 
= \frac{1}{4\pi^2 k_F}\int dq \left[ \frac{1}{2\tilde{\chi}_d m^{*}} \left( l^2 - \frac{4l}{\tilde{\phi}} \right) +O(q) \right]
\label{eq:expansion}
\end{align}
for $qd\ll 1$.
It gives a good guideline for understanding the stable pairing symmetry. The quantity $l^2-4l/\tilde{\phi}$ is negative for $\tilde{\phi}=2$ and $l=+1$, which explains negative $\Delta\lambda_{\phi,m}^{(l)}$ at $\nu = \frac{1}{2}+\frac{1}{2}$. It also nicely dictates the ordering of $\Delta\lambda_{\phi,m}^{(l)}$ at low frequencies, while higher order corrections should be considered if $l^2-4l/\tilde{\phi}=0$. For example, at $\nu=\frac{1}{4}+\frac{1}{4}$, $l=+1$ gives $l^2-4l/\tilde{\phi}=0$, but still the $l=+1$ state is favored. 

The small-$q$ expansion \eqref{eq:expansion} moreover reveals the mechanism of stabilizing the $l=+1$ state. The $l^2$ term originates from the current-current interaction and the $4l/\tilde{\phi}$ term from the density-current interaction. Both are mediated by the out-of-phase gauge fluctuation. Since the current-current interaction is isotropic, it favors the $l=0$ state and increases the energy of paired states with higher angular momentum. In contrast, the density-current interaction can be attractive or repulsive depending on the direction of the external magnetic field and the pairing symmetry. If it is attractive and exceeds the repulsion for the $l\neq 0$ states, there is a chance of pairing with finite orbital angular momentum. This occurs only for $l=+1$ and $\tilde{\phi}\leq 4$ (provided $\tilde{\phi}>0$), which explains the stable $l=+1$ state. 

The $l=+1$ state of CFs has the opposite angular momentum to the fluxes attached to electrons. This is seen from the electron wave function [Eq.~\eqref{eq:wave}]. 
For small distances, it has a form 
\begin{equation}
\Psi(\{z\}, \{w\}) \approx \prod_{i<j} (z_i-z_j)^{\tilde{\phi}} \prod_{i'<j'} (w_{i'}-w_{j'})^{\tilde{\phi}}  \cdot \det\left( \frac{1}{z_i-w_j} \right),
\end{equation}
which shows the opposite angular momenta for the fluxes and interlayer pairing. 

Our finding of the interlayer paired state with $l=+1$ at large layer spacing is consistent with a preceding study \cite{Morinari}, which estimated the pairing symmetry within the BCS theory. 
The properties of this $l=+1$ state are studied also in Ref.~\cite{Kim} without energetics. 
On the other hand, numerical studies of finite size quantum Hall bilayers on a sphere seem to infer a paired CF phase of the $l=-1$ interlayer paired state at $\nu=\frac{1}{2}+\frac{1}{2}$ \cite{numerics,trial}. 
This $l=-1$ state was found to be an exciton condensate in a very recent paper \cite{inti-senthil}, which preserves the particle-hole symmetry of half-filled Landau levels. 
Here, we focus on the time-reversal symmetry breaking due to the external magnetic field, instead of the particle-hole symmetry, only present in the case of $\nu=\frac{1}{2}+\frac{1}{2}$. 
The origin of the discrepancy in the stable pairing channel is presently unclear.

\textit{Conclusion.}---%
We have studied the pairing symmetry of interlayer paired states in quantum Hall bilayers by taking into account of the density and current fluctuations of CFs, and have found the $l=+1$ pairing is energetically favored at the filling fraction $\nu=\frac{1}{2}+\frac{1}{2}$ and $\frac{1}{4}+\frac{1}{4}$. The Chern-Simons term couples the density and current fluctuations, which breaks the time-reversal symmetry to lift the degeneracy of $\pm l$ states.

\textit{Acknowledgment}.---%
We thank A. V. Chubukov, T. Senthil, and I. Sodemann for valuable discussions. 
This work is supported by the U.S.\ DOE Office of Basic Energy Sciences, Division of Materials Sciences and Engineering under Award No.\ DE-SC0010526.


\onecolumngrid
\noindent

\begin{center}
{\large\bf Supplemental Material}
\end{center}

\setcounter{equation}{0}
\setcounter{figure}{0}
\def\theequation{S\arabic{equation}}
\def\thefigure{S\arabic{figure}}

\section{RPA calculation}

We derive the Chern-Simons gauge field propagator with the random-phase approximation (RPA). 
The model we consider is already given in the main text. Here we repeat for convenience: 
\begin{equation}
Z=\int \prod_s D\psi_s^\dagger D\psi_s D\bm{a}^{(s)} Da_0^{(s)} e^{-S},
\end{equation}
where the action $S$ is 
\begin{equation}
S=\int_0^\beta d\tau \int d^2 r \mathcal{L} (\bm{r},\tau),
\end{equation}
and the Lagrangian density $\mathcal{L}$ is
\begin{align}
\mathcal{L}(\bm{r},\tau) 
=& \sum_{s} \bigg\{ \psi_s^{\dagger}(\bm{r},\tau) \left[\partial_{\tau}+ia_{0}^{(s)}(\bm{r},\tau) \right] \psi_s(\bm{r},\tau) 
+\frac{1}{2m^{*}}\psi_s^{\dagger}(\bm{r},\tau) \left[ -i\nabla-\bm{a}^{(s)}(\bm{r},\tau) +e\bm{A}(\bm{r}) \right]^{2}\psi_s(\bm{r},\tau) \bigg\} \notag \\
& -\sum_{ss'}\frac{i}{2\pi}K_{ss'}^{-1}a_{0}^{(s)}(\bm{r},\tau) \hat{z}\cdot [\nabla\times \bm{a}^{(s')}(\bm{r},\tau)] 
+\frac{1}{2}\sum_{ss'}\int d^{2}r'\delta\rho_s(\bm{r},\tau)V_{ss'}(\bm{r}-\bm{r}')\delta\rho_{s'}(\bm{r}',\tau). 
\label{eq:lagrangian}
\end{align}
We assume the Coulomb (transverse) gauge for the Chern-Simons gauge field; $\nabla\cdot\bm{a}^{(s)}=0$. 
The long-range Coulomb interaction 
\begin{equation}
V_{ss'}(r)=\frac{e^{2}}{\varepsilon\sqrt{r^{2}+(1-\delta_{s,s'})d^{2}}}
\end{equation}
acts on composite fermions and its Fourier transform is 
\begin{equation}
V_{ss'}(q)=\frac{2\pi e^{2}}{\varepsilon q}e^{-qd(1-\delta_{ss'})}.
\end{equation}
The $K$-matrix is 
\begin{equation}
K_{ss'}=\begin{pmatrix}
\tilde{\phi} & 0\\
0 & \tilde{\phi}
\end{pmatrix},
\end{equation}
with the integer $\tilde{\phi}$ corresponding to the number of fluxes attached to an electron. 
The composite fermion density fluctuation is given by
\begin{equation}
\delta\rho_{s}(\bm{r},\tau)=\psi_{s}^{\dagger}(\bm{r},\tau)\psi_{s}(\bm{r},\tau)-n_e,
\end{equation}
where $n_e$ is the electron density. 

Since we assume the Coulomb gauge for the Chern-Simons gauge field, the transverse part of the gauge field $a_1$ can be written as 
\begin{equation}
a_{1}^{(s)}(\bm{q},\tau)=\hat{z}\cdot\left[\hat{\bm{q}}\times\bm{a}^{(s)}(\bm{q},\tau)\right], 
\end{equation}
or inversely
\begin{equation}
\bm{a}^{(s)}(\bm{q},\tau)=a_{1}^{(s)}(\bm{q},\tau)(\hat{z}\times\hat{\bm{q}}). 
\end{equation}

From Eq.~\eqref{eq:lagrangian}, the Green's function for the composite fermions is 
\begin{equation}
G (\bm{k},i\epsilon_n) = \frac{1}{i\epsilon_n -\epsilon_{\bm{k}}}, 
\end{equation}
the bare gauge propagator is 
\begin{equation}
D^{(0)}_{s\mu,s'\nu} (\bm{q},i\omega_m)^{-1} = 
\begin{pmatrix}
0 & \frac{q}{2\pi\phi}\\
\frac{q}{2\pi\phi} & -\frac{q^{2}V_{ss'}}{(2\pi\phi)^{2}}
\end{pmatrix}_{\mu\nu},
\end{equation}
and the vertices are diagrammatically given by 
\begin{gather}
\label{eq:vertex0}
\parbox[c]{1.2cm}{\includegraphics[width=1.2cm]{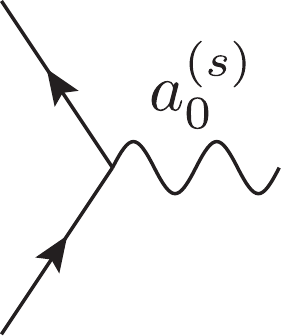}}
=-i\psi_{s}^{\dagger}(\bm{k}+\bm{q},i\epsilon_{n}+i\omega_{m})a_{0}^{(s)}(\bm{q},i\omega_{m})\psi_{s}(\bm{k},i\epsilon_{n}), \\
\label{eq:vertex1}
\parbox[c]{1.2cm}{\includegraphics[width=1.2cm]{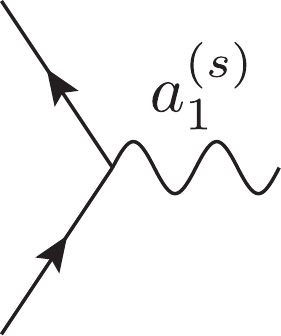}}
=\dfrac{1}{m^{*}}\psi_{s}^{\dagger}(\bm{k}+\bm{q},i\epsilon_{n}+i\omega_{m})\left[\hat{z}\cdot(\hat{\bm{q}}\times\bm{k})\right]a_{1}^{(s)}(\bm{q},i\omega_{m})\psi_{s}(\bm{k},i\epsilon_{n}), \\
\label{eq:vertex11}
\parbox[c]{1.5cm}{\includegraphics[width=1.5cm]{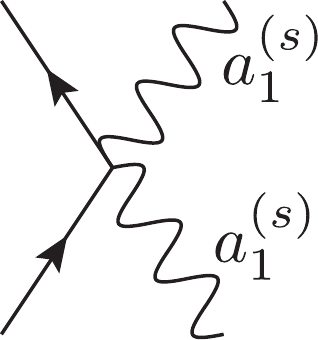}}
=\dfrac{1}{2m^{*}}\psi_{s}^{\dagger}(\bm{k}+\bm{q}-\bm{q}',i\epsilon_{n}+i\omega_{m}-i\omega_{m}')(\hat{\bm{q}}\cdot\hat{\bm{q}}')a_{1}^{(s)}(\bm{q},i\omega_{m})a_{1}^{(s)}(-\bm{q}',-i\omega_{m}')\psi_{s}(\bm{k},i\epsilon_{n}). 
\end{gather}

\begin{figure}
\includegraphics[width=12cm]{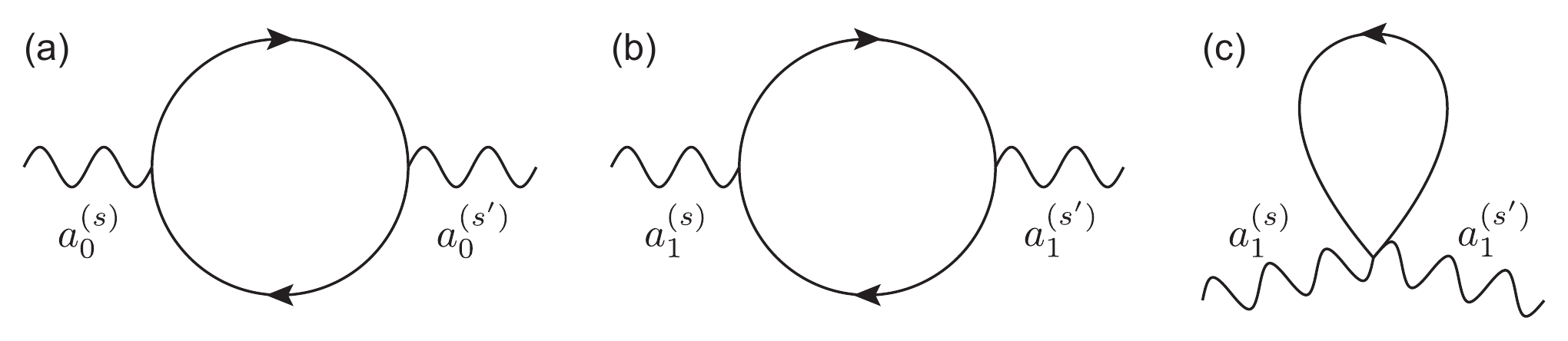}

\caption{Polarization diagrams: (a) $\Pi_{00}$, (b) $\Pi_{11,p}$, and (c) $\Pi_{11,d}$.}
\label{fig:pol}
\end{figure}

Now we calculate one-loop diagrams for the gauge propagator (Fig.~\ref{fig:pol})
\begin{align}
\Pi_{00}^{ss'}(q,i\omega_{m}) & =\langle a_{0}^{(s)}(\bm{q},i\omega_{m})a_{0}^{(s')}(-\bm{q},-i\omega_{m})\rangle=\Pi_{00}(q,i\omega_{m})\delta_{ss'},\\
\Pi_{11}^{ss'}(q,i\omega_{m}) & =\Pi_{11,p}^{ss'}(\bm{q},i\omega_{m})+\Pi_{11,d}^{ss'}(\bm{q},i\omega_{m}),\\
\Pi_{11,p}^{ss'}(q,i\omega_{m}) & =\langle a_{1}^{(s)}(\bm{q},i\omega_{m})a_{1}^{(s)}(-\bm{q},-i\omega_{m})\rangle_{\text{paramag}}=\Pi_{11,p}(q,i\omega_{m})\delta_{ss'},\\
\Pi_{11,d}^{ss'}(q,i\omega_{m}) & =\langle a_{1}^{(s)}(\bm{q},i\omega_{m})a_{1}^{(s)}(-\bm{q},-i\omega_{m})\rangle_{\text{diamag}}=\Pi_{11,d}(q,i\omega_{m})\delta_{ss'},
\end{align}
and the other components vanish. 
Each diagram is calculated as follows:
\begin{align}
\Pi_{00}(q,i\omega_{m}) & =(-1)(-i)^{2}T\sum_{\epsilon_{n}}\int \frac{d^2k}{(2\pi)^2}G(\bm{k}+\bm{q},i\epsilon_{n}+i\omega_{m})G(\bm{k},i\epsilon_{n})\nonumber \\
 & =-\int \frac{d^2k}{(2\pi)^2}\frac{f(\epsilon_{\bm{k}+\bm{q}})-f(\epsilon_{\bm{k}})}{i\omega_{m}-\epsilon_{\bm{k}+\bm{q}}+\epsilon_{\bm{k}}}\nonumber \\
 & =F_{1}(q,i\omega_{m})+F_{1}(q,-i\omega_{m}),
\end{align}
\begin{align}
\Pi_{11,p}(q,i\omega_{m}) & =(-1)T\sum_{\epsilon_{n}}\int \frac{d^2k}{(2\pi)^2}\frac{\hat{z}\cdot\hat{\bm{q}}\times\left(\bm{k}+\frac{\bm{q}}{2}\right)}{m^{*}}\frac{\hat{z}\cdot(-\hat{\bm{q}})\times\left(\bm{k}+\frac{\bm{q}}{2}\right)}{m^{*}}G(\bm{k}+\bm{q},i\epsilon_{n}+i\omega_{m})G(\bm{k},i\epsilon_{n})\nonumber \\
 & =-\int \frac{d^2k}{(2\pi)^2}\left(\frac{\hat{\bm{q}}\times\bm{k}}{m^{*}}\right)^{2}\frac{f(\epsilon_{\bm{k}+\bm{q}})-f(\epsilon_{\bm{k}})}{i\omega_{m}-\epsilon_{\bm{k}+\bm{q}}+\epsilon_{\bm{k}}}\nonumber \\
 & =F_{2}(q,i\omega_{m})+F_{2}(q,-i\omega_{m}),
\end{align}
\begin{align}
\Pi_{11,d}(q,i\omega_{m}) & =2(-1)T\sum_{\epsilon_{n}}\int \frac{d^2k}{(2\pi)^2}\frac{1}{2m^{*}}\left[-\hat{\bm{q}}^{2}G(\bm{k},i\epsilon_{n})\right]\nonumber \\
 & =\frac{1}{m^{*}}T\sum_{\epsilon_{n}}\int \frac{d^2k}{(2\pi)^2}G(\bm{k},i\epsilon_{n})\nonumber \\
 & =\frac{n_e}{m^{*}}=\frac{\epsilon_{F}}{2\pi}.
\end{align}
Here the functions $F_1(\bm{q},i\omega_m)$ and $F_2(\bm{q},i\omega_m)$ are defined by \begin{gather}
F_{1}(q,i\omega_{m})  =\int \frac{d^2k}{(2\pi)^2}\frac{f(\epsilon_{\bm{k}})}{i\omega_{m}-\epsilon_{\bm{k}+\bm{q}}+\epsilon_{\bm{k}}},\\
F_{2}(q,i\omega_{m})  =\int \frac{d^2k}{(2\pi)^2}\left(\frac{\hat{\bm{q}}\times\bm{k}}{m^{*}}\right)^{2}\frac{f(\epsilon_{\bm{k}})}{i\omega_{m}-\epsilon_{\bm{k}+\bm{q}}+\epsilon_{\bm{k}}}.
\end{gather}

At $T=0$, those functions are calculated analytically \cite{s-Bonesteel3}. We write
\begin{gather}
F_{1}(q,i\omega_{m}) =2m^{*}f_{1}\left(\frac{2q}{k_{F}},\frac{i\omega_{m}}{\epsilon_{F}}-\frac{q^{2}}{k_{F}^{2}}\right),\\
F_{2}(q,i\omega_{m}) =4\epsilon_{F}f_{2}\left(\frac{2q}{k_{F}},\frac{i\omega_{m}}{\epsilon_{F}}-\frac{q^{2}}{k_{F}^{2}}\right),
\end{gather}
where the functions $f_1$ and $f_2$ are
\begin{gather}
f_{1}(y,z)  =\frac{1}{(2\pi)^{2}}\int_{0}^{1}dx\int_{0}^{2\pi}d\theta\frac{1}{z-xy\cos\theta},\\
f_{2}(y,z)  =\frac{1}{(2\pi)^{2}}\int_{0}^{1}dx\int_{0}^{2\pi}d\theta\frac{x^{2}\sin^{2}\theta}{z-xy\cos\theta}.
\end{gather}
One can perform the $\theta$ integrations by contour integrals on the
complex plane, keeping in mind the analytic continuation $i\omega_{m}\to\omega+i\delta$.
Then we obtain
\begin{gather}
f_{1}(y,z) =\frac{1}{2\pi y}\frac{z}{y}\left[1-\left(1-\frac{y^{2}}{z^{2}}\right)^{1/2}\right],\\
f_{2}(y,z) =\frac{1}{4\pi y}\frac{z}{y}\left\{ 1-\frac{2}{3}\frac{z^{2}}{y^{2}}\left[1-\left(1-\frac{y^{2}}{z^{2}}\right)^{3/2}\right]\right\} .
\end{gather}
Note that both $f_{1}(y,z)$ and $f_{2}(y,z)$ have branch cuts between
$z=+y$ and $z=-y$. 

\begin{figure}
\includegraphics[width=0.7\hsize]{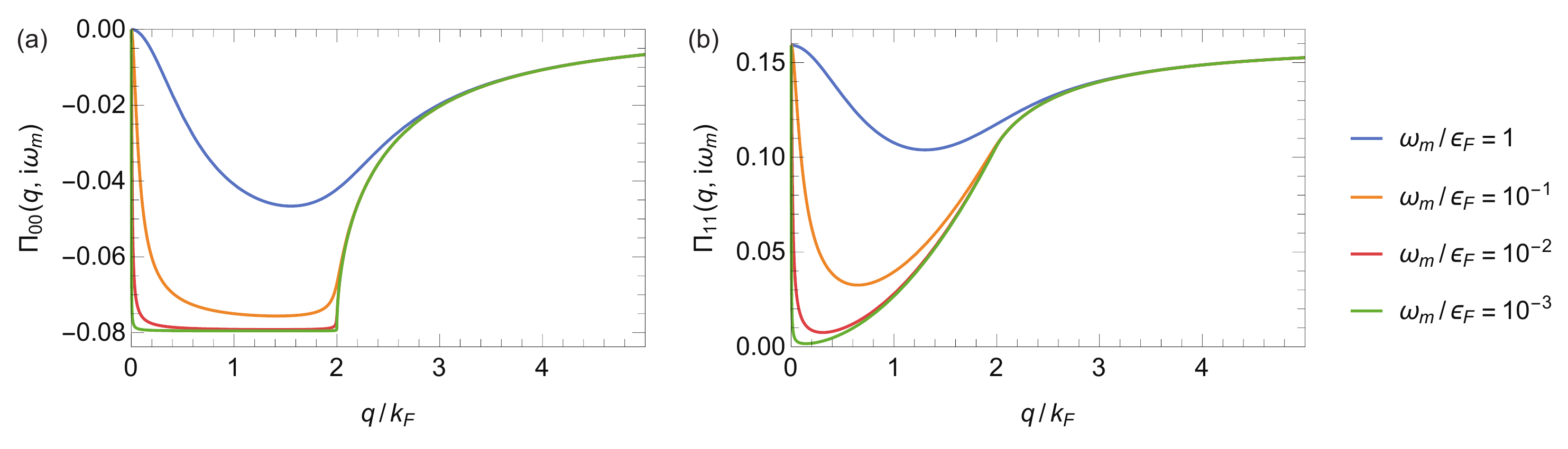}
\caption{%
$\Pi_{00}(q,i\omega_m)$ and $\Pi_{11}(q,i\omega_m)$. The polarization functions are real before the analytic continuation to real frequencies.  }
\label{fig:pol_im}
\end{figure}

\begin{figure}
\includegraphics[width=0.6\hsize]{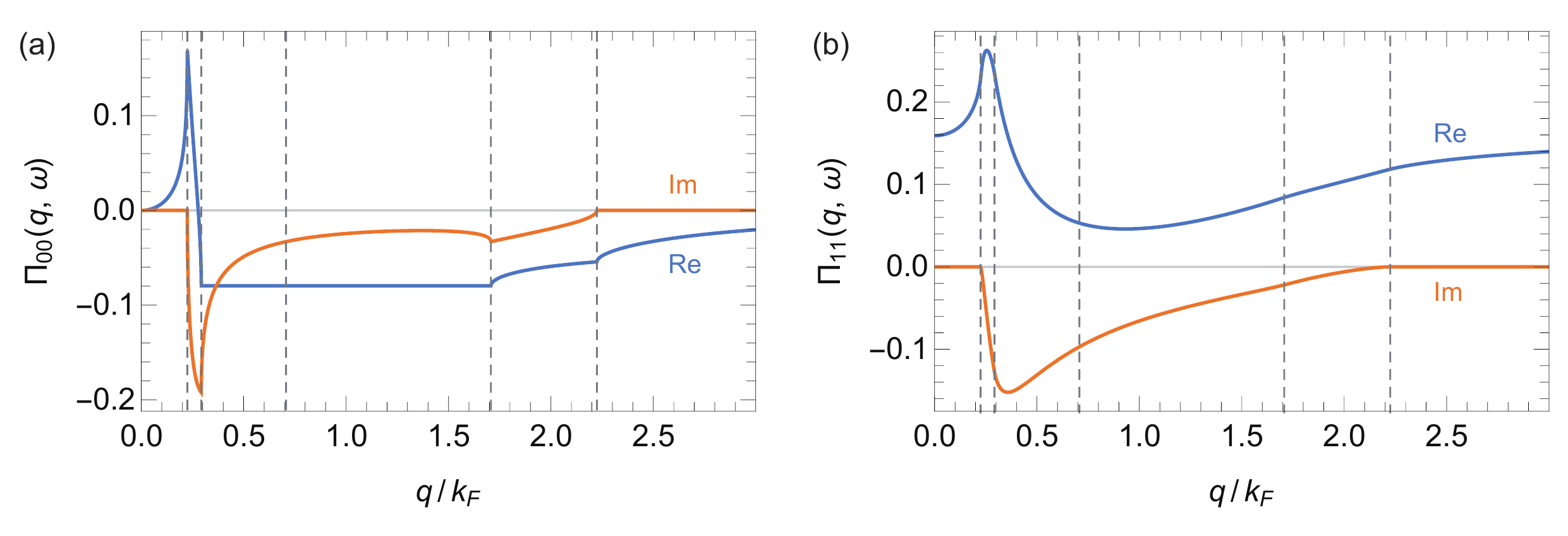}
\caption{%
$\Pi_{00}(q,\omega)$ and $\Pi_{11}(q,\omega)$ with $\omega=0.5\epsilon_{F}$.
Blue and orange lines are real and imaginary parts, respectively.
Vertical dashed lines represent characteristic momenta, where analytic expressions
change. }
\label{fig:pol_re}
\end{figure}

Now we have the analytic expressions of one-loop polarization functions $\Pi_{00}$ and $\Pi_{11}$. Figures~\ref{fig:pol_im} and \ref{fig:pol_re} show the polarization functions before and after analytic continuation $i\omega \to \omega+i\delta$. After analytic continuation, analytic expressions change at 
\begin{align}
q(\omega) = k_F \sqrt{\frac{\omega}{\epsilon_F}}, \ 
k_F \sqrt{2-\frac{\omega}{\epsilon_F}\pm\sqrt{4-\frac{4\omega}{\epsilon_F}}}, \ 
k_F \sqrt{2+\frac{\omega}{\epsilon_F}\pm\sqrt{4+\frac{4\omega}{\epsilon_F}}}. 
\end{align}
 
It is useful to see some approximate forms of the polarization functions:
\newline\noindent
(a) 
$|\omega_m|/\epsilon_{F}\gg (q/k_{F})^2$:  
\begin{align}
\Pi_{00}(q,i\omega_m) & \approx -\frac{\epsilon_{F}}{2\pi}\frac{q^{2}}{\omega_m^{2}},\\
\Pi_{11,p}(q,i\omega_m) & \approx \frac{\epsilon_{F}^{2}}{4\pi m^{*}}\frac{q^{2}}{\omega_m^{2}},\\
\Pi_{11}(q,i\omega_m) & \approx\frac{\epsilon_{F}}{2\pi}\left(1+\frac{\epsilon_{F}}{2m^{*}}\frac{q^{2}}{\omega_m^{2}}\right)\approx\frac{\epsilon_{F}}{2\pi}.
\end{align}
\newline\noindent
(b) 
$|\omega_m|/\epsilon_{F}\ll (q/k_{F})^2 \ll 1$:
\begin{align}
\Pi_{00}(q,i\omega_m) & \approx-\frac{m^{*}}{2\pi},\\
\Pi_{11,p}(q,i\omega_m) & \approx-\frac{\epsilon_{F}}{2\pi}-\frac{m^{*}}{2\pi}\frac{\omega_m^{2}}{q^{2}}+\frac{1}{24\pi m^{*}}q^{2}+\frac{k_{F}}{2\pi}\frac{|\omega_m|}{q},\\
\Pi_{11}(q,i\omega_m) & \approx\chi_{d}q^{2}-\frac{m^{*}}{2\pi}\frac{\omega_m^{2}}{q^{2}}+\frac{k_{F}}{2\pi}\frac{|\omega_m|}{q},
\end{align}
where $\chi_d$ is the diamagnetic susceptibility 
\begin{equation}
\chi_{d}=\frac{1}{24\pi m^{*}}.
\end{equation}

\begin{figure}
\includegraphics[width=10cm]{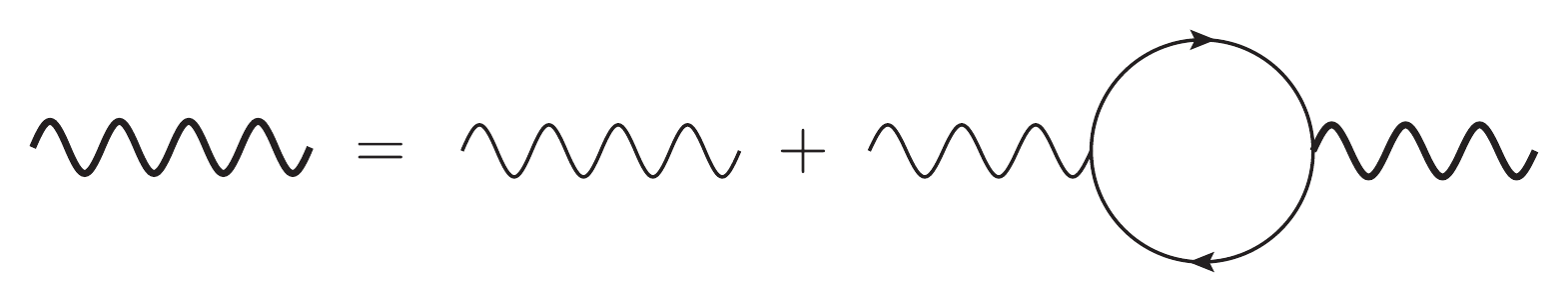}
\caption{RPA gauge field propagator $D_{s\mu,s'\nu}(q,i\omega_{m})$ (bold wavy line). Thin wavy lines represent the bare gauge field propagator $D^{(0)}_{s\mu,s'\nu}(q,i\omega_{m})$.}
\label{fig:rpa}
\end{figure}

The effective action for the gauge field is given by 
\begin{align}
S_{\text{eff}}[a] & =\frac{1}{2}T\sum_{\omega_{m}}\int \frac{d^2q}{(2\pi)^2}\sum_{s,s'=1,2}\sum_{\mu,\nu=0,1}a_{\mu}^{(s)}(\bm{q},i\omega_{m})D_{s\mu,s'\nu}^{-1}(q,i\omega_{m})a_{\nu}^{(s')}(-\bm{q},-i\omega_{m}),\label{eq:Seff}
\end{align}
where $D(q,i\omega_m)$ is the RPA gauge field propagator, calculated by 
\begin{equation}
D_{s\mu,s'\nu} (q,i\omega_m)^{-1} = D^{(0)}_{s\mu,s'\nu} (q,i\omega_m)^{-1} - \Pi_{\mu\nu}^{ss'} (q,i\omega_m),
\end{equation}
or 
\begin{equation}
D_{s\mu,s'\nu} (q,i\omega_m) = 
D^{(0)}_{s\mu,s'\nu} (q,i\omega_m) +
D^{(0)}_{s\mu,s'\nu} (q,i\omega_m) \Pi_{\mu\nu}^{ss'} (q,i\omega_m) D_{s\mu,s'\nu} (q,i\omega_m). 
\end{equation}
The diagrammatic expression is given in Fig.~\ref{fig:rpa}. 
The nonzero components are 
\begin{gather}
D_{s0,s'0}^{-1}(q,i\omega_{m})  =-\Pi_{00}(q,i\omega_{m})\delta_{ss'},\\
D_{s1,s'1}^{-1}(q,i\omega_{m})  =-\Pi_{11}(q,i\omega_{m})\delta_{ss'}-\frac{1}{(2\pi\tilde{\phi})^{2}}q^{2}V_{ss'}(q),\\
D_{s0,s'1}^{-1}(q,i\omega_{m})  =D_{s1,s'0}^{-1}(q,i\omega_{m})=\frac{q}{2\pi}K_{ss'}^{-1}.
\end{gather}
It is convenient to use the in-phase and out-of-phase basis for the Chern-Simons gauge 
\begin{equation}
a_\mu^{(\pm)} \equiv \frac{1}{\sqrt{2}} (a_\mu^{(1)}\pm a_\mu^{(2)}). 
\end{equation}
Then $D^{-1}(\bm{q},i\omega_{m})$ becomes 
\begin{align}
& D^{-1}(q,i\omega_{m}) \notag \\
=&
\begin{pmatrix}
-\Pi_{00}(q,i\omega_{m}) & \frac{q}{2\pi\tilde{\phi}}\\
\frac{q}{2\pi\tilde{\phi}} & -\Pi_{11}(q,i\omega_{m})-\frac{q^{2}V_{11}(q)}{(2\pi\tilde{\phi})^{2}}-\frac{q^{2}V_{12}(q)}{(2\pi\tilde{\phi})^{2}}\\
 &  & -\Pi_{00}(q,i\omega_{m}) & \frac{q}{2\pi\tilde{\phi}}\\
 &  & \frac{q}{2\pi\tilde{\phi}} & -\Pi_{11}(q,i\omega_{m})-\frac{q^{2}V_{11}(q)}{(2\pi\tilde{\phi})^{2}}+\frac{q^{2}V_{12}(q)}{(2\pi\tilde{\phi})^{2}}
\end{pmatrix} \notag \\
\equiv &
\begin{pmatrix}
D_{+,\mu\nu}^{-1}(q,i\omega_{m})\\
 & D_{-,\mu\nu}^{-1}(q,i\omega_{m})
\end{pmatrix}.
\end{align}
This shows that the in-phase ($+$) and out-of-phase ($-$) modes are decoupled. 

The determinants of the two matrices $D_{\pm,\mu\nu}^{-1}(q,i\omega_{m})$ are obtained as 
\begin{equation}
\det D_{\pm}^{-1}(q,i\omega_{m})=\Pi_{00}(q,i\omega_{m})\left[\Pi_{11}(q,i\omega_{m})+\frac{q^{2}V_{11}(q)}{(2\pi\tilde{\phi})^{2}}\pm\frac{q^{2}V_{12}(q)}{(2\pi\tilde{\phi})^{2}}\right]-\frac{q^{2}}{(2\pi\tilde{\phi})^{2}}.
\end{equation}
Their zeros correspond to collective modes for the in-phase and out-of-phase fluctuations, respectively. 
The matrices $D_{\pm}^{-1}(q,i\omega_{m})$
can be easily inverted to obtain
\begin{align}
\label{eq:propagator}
D_{\pm}(q,i\omega_{m}) & 
=-\frac{1}{\det D_{\pm}^{-1}(q,i\omega_{m})}
\begin{pmatrix}
\Pi_{11}(q,i\omega_{m})+\frac{q^{2}V_{11}(q)}{(2\pi\tilde{\phi})^{2}}\pm\frac{q^{2}V_{12}(q)}{(2\pi\tilde{\phi})^{2}} & \frac{q}{2\pi\tilde{\phi}}\\
\frac{q}{2\pi\tilde{\phi}} & \Pi_{00}(q,i\omega_{m})
\end{pmatrix}.
\end{align}
Using the relations
\begin{align}
D_{\pm,\mu\nu}(q,i\omega_{m})=\langle a_{\mu}^{(\pm)}(q,i\omega_{m}) a_{\nu}^{(\pm)}(-q,-i\omega_{m}) \rangle , \\
D_{s\mu,s'\nu}(q,i\omega_{m})=\langle a_{\mu}^{(s)}(q,i\omega_{m}) a_{\nu}^{(s')}(-q,-i\omega_{m}) \rangle ,
\end{align}
$D_{s\mu,s'\nu}(q,i\omega_{m})$ written as 
\begin{align}
D_{s\mu,s'\nu}(q,i\omega_{m}) & =
\begin{cases}
\dfrac{1}{2}\left[D_{+,\mu\nu}(q,i\omega_{m})+D_{-,\mu\nu}(q,i\omega_{m})\right] & (s=s'\text{; intralayer}), \vspace{3pt} \\ 
\dfrac{1}{2}\left[D_{+,\mu\nu}(q,i\omega_{m})-D_{-,\mu\nu}(q,i\omega_{m})\right] & (s\neq s'\text{; interlayer}). 
\end{cases}
\label{eq:dpm}
\end{align}

For $|\omega_m|/\epsilon_{F}\ll (q/k_{F})^2 \ll1$ and $q\ll d^{-1}$, $D_{\pm} (q,i\omega_m)$ is approximated as 
\begin{gather}
D_{+,\mu\nu}(q,i\omega_m)  \approx
\frac{1}{\frac{e^2}{\pi\varepsilon\tilde{\phi}^{2}}q + \frac{k_{F}}{2\pi}\frac{|\omega_m|}{q}}
\begin{pmatrix}
\frac{2\pi}{m^*} \left(\frac{e^2}{\pi\varepsilon\tilde{\phi}^{2}}q+\frac{k_{F}}{2\pi}\frac{|\omega_m|}{q}\right) & \frac{q}{m^* \tilde{\phi}}\\
\frac{q}{m^* \tilde{\phi}} & -1
\end{pmatrix},
\label{eq:dp}\\
D_{-,\mu\nu}(q,i\omega_m)  \approx
\frac{1}{\tilde{\chi}_{d}q^{2}+\frac{k_{F}}{2\pi}\frac{|\omega_m|}{q}}
\begin{pmatrix}
\frac{2\pi}{m^{*}} \left[ \left(\chi_d+\frac{e^2 d}{2\pi\varepsilon\tilde{\phi}^2}\right)q^{2}+\frac{k_{F}}{2\pi}\frac{|\omega_m|}{q} \right] & \frac{q}{m^* \tilde{\phi}}\\
\frac{q}{m^* \tilde{\phi}} & -1
\end{pmatrix},
\label{eq:dm}
\end{gather}
where $\tilde{\chi}_{d}$ is defined as 
\begin{equation}
\tilde{\chi}_{d}=\chi_d+\frac{e^{2}d}{2\pi\varepsilon\tilde{\phi}^{2}}+\frac{1}{2\pi m^{*}\tilde{\phi}^{2}}.
\end{equation}
We note 
\begin{align}
V_{11}(q)\pm V_{12}(q) & =\frac{2\pi e^{2}}{\epsilon q}(1\pm e^{-qd})\simeq
\begin{cases}
\dfrac{4\pi e^{2}}{\varepsilon q} & (+)\\
\dfrac{2\pi e^{2}d}{\varepsilon} & (-)
\end{cases}
\end{align} 
for $q\ll d^{-1}$.
Here we can observe that $D_{-,11}(q,i\omega_m)$ is the most singular term for small $q$ in $D_{\pm,\mu\nu}(q,i\omega_m)$, followed by $D_{+,11}(q,i\omega_m)$, $D_{-,01}(q,i\omega_m)$ and $D_{-,10}(q,i\omega_m)$.

For $|\omega_m|/\epsilon_{F} \gg (q/k_{F})^2$ and $q\ll d^{-1}$, the approximate forms of $D_{\pm} (q,i\omega_m)$  are 
\begin{gather}
D_{+,\mu\nu}(q,i\omega_m)  \approx
\frac{1}{\left(\frac{\epsilon_F}{2\pi}\right)^2 \frac{q^2}{\omega_m^2}+ \frac{q^2}{(2\pi\tilde{\phi})^2}}
\begin{pmatrix}
\frac{\epsilon_F}{2\pi}+\frac{e^2 q}{\pi\varepsilon\tilde{\phi}^2} & -\frac{q}{2\pi\tilde{\phi}}\\
-\frac{q}{2\pi\tilde{\phi}} & -\frac{\epsilon_F}{2\pi}\frac{q^2}{\omega_m^2}
\end{pmatrix},
\label{eq:dp_h}
\\
D_{-,\mu\nu}(q,i\omega_m)  \approx
\frac{1}{\left(\frac{\epsilon_F}{2\pi}\right)^2 \frac{q^2}{\omega_m^2}+ \frac{q^2}{(2\pi\tilde{\phi})^2}}
\begin{pmatrix}
\frac{\epsilon_F}{2\pi}+\frac{e^2 d q^2}{2 \pi\varepsilon\tilde{\phi}^2} & -\frac{q}{2\pi\tilde{\phi}}\\
-\frac{q}{2\pi\tilde{\phi}} & -\frac{\epsilon_F}{2\pi}\frac{q^2}{\omega_m^2}
\end{pmatrix}. 
\label{eq:dm_h}
\end{gather}

\section{Effective interaction}

\begin{figure}
\includegraphics[width=0.6\hsize]{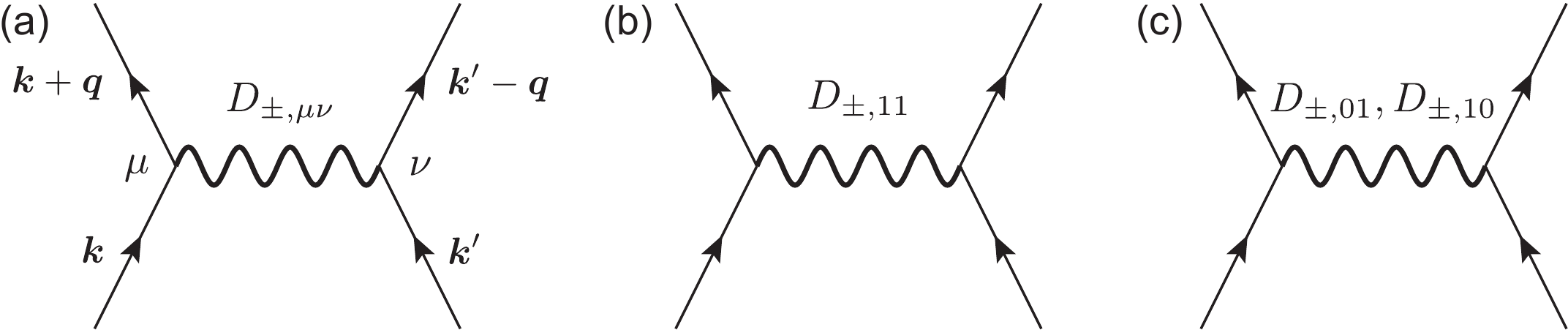}
\caption{
(a) Effective interaction between composite fermions. 
(b) Effective interaction mediated by the current-current gauge propagator. 
(c) Effective interaction via the Chern-Simons term. 
}
\label{fig:eff1}
\end{figure}

The effective interaction acting on composite fermions is mediated by the Chern-Simons gauge field. It is diagrammatically given in Fig.~\ref{fig:eff1}(a), which is written as 
\begin{align}
\label{eq:eff_int}
\mathcal{V}=\frac{1}{2}\sum_{s_1 s_2 s_3 s_4} V_{s_1 s_2 s_3 s_4}^{\text{eff}}(\bm{k},\bm{k}',\bm{q},i\omega_m)  \psi_{s_1}^{\dagger}(\bm{k}+\bm{q},i\epsilon_n+i\omega_m) \psi_{s_2}^\dagger (\bm{k}'-\bm{q},i\epsilon'_n-i\omega_m) 
\psi_{s_3}(\bm{k}',i\epsilon'_n) \psi_{s_4} (\bm{k},i\epsilon_n), 
\end{align}
where the matrix element is given by
\begin{align}
V_{s_{1}s_{2}s_{3}s_{4}}^{\text{eff}}(\bm{k},\bm{k}',\bm{q},i\omega_{m})
= \sum_{\mu,\nu=0,1}M_{\mu\nu}(\bm{k},\bm{k}',\hat{\bm{q}}) \bigl[D_{+,\mu\nu}(q,i\omega_{m})(\sigma_{0})_{s_{1}s_{4}}(\sigma_{0})_{s_{2}s_{3}} +D_{-,\mu\nu}(q,i\omega_{m})(\sigma_{3})_{s_{1}s_{4}}(\sigma_{3})_{s_{2}s_{3}} \bigr]. 
\label{eq:effective}
\end{align}
The matrix $M_{\mu\nu}(\bm{k},\bm{k}',\hat{\bm{q}})$ reflects the forms of the vertices \eqref{eq:vertex0} and \eqref{eq:vertex1} and becomes
\begin{equation}
\label{eq:matrix}
M_{\mu\nu}(\bm{k},\bm{k}',\hat{\bm{q}})=\frac{1}{2}\begin{pmatrix}1 & -i\frac{\hat{z}\cdot(\hat{\bm{q}}\times\bm{k}')}{m^*}\\
i\frac{\hat{z}\cdot(\hat{\bm{q}}\times\bm{k})}{m^*} & \frac{(\hat{\bm{q}}\times\bm{k})\cdot(\hat{\bm{q}}\times\bm{k}')}{m^{*2}}
\end{pmatrix}_{\mu\nu}. 
\end{equation}

By calculating Eq.~\eqref{eq:effective}, one finds only the current-current interaction mediated by $D_{\pm,11}$ [Fig.~\ref{fig:eff1}(b)] has a singularity at small $q$. 
When we consider the interlayer interaction, the current-current contribution is given by 
\begin{align}
-\frac{(\hat{\bm{q}}\times\bm{k})\cdot(\hat{\bm{q}}\times\bm{k}')}{2m^{*2}} \left[ D_{-,11}(q,i\omega_m) - D_{+,11}(q,i\omega_m) \right],
\end{align}
and it is attractive in the Cooper channel $\bm{k}' =-\bm{k}$. 
Only this contribution is considered in Ref.~\cite{s-Bonesteel2}, since the singularity at small $q$ is important to analyze the instability for the formation of a paired state. 
Still there are other contributions in the effective interaction, and those terms turn out to play an important role for determining the pairing symmetry. 
For example, time-reversal symmetry breaking is not captured within that approximation, and the effect from the Chern-Simons term needs to be included. 

The importance of the off-diagonal terms that reflect the Chern-Simons term is seen by the following argument: First consider the operators
\begin{align}
\mathcal{O}_{+}^\dagger (\bm{k}) & =(k_{x}+ik_{y})(c_{\bm{k}\uparrow}^{\dagger}c_{-\bm{k}\downarrow}^{\dagger}+c_{\bm{k}\downarrow}^{\dagger}c_{-\bm{k}\uparrow}^{\dagger}),\\
\mathcal{O}_{-}^\dagger (\bm{k}) & =(k_{x}-ik_{y})(c_{\bm{k}\uparrow}^{\dagger}c_{-\bm{k}\downarrow}^{\dagger}+c_{\bm{k}\downarrow}^{\dagger}c_{-\bm{k}\uparrow}^{\dagger}),
\end{align}
where $\mathcal{O}_{\pm}$ corresponds to a creation of an interlayer paired states with a relative angular momentum $l=\pm 1$. 
They are equivalent to spin-triplet $(p_{x}\pm ip_{y})$-wave pairings, with the spins $\uparrow$, $\downarrow$ considered as layer indices. 
Then we calculate a quantity 
\begin{align}
 & \mathcal{O}_{+}^{\dagger}(\bm{k})\mathcal{O}_{+}(\bm{k}')-\mathcal{O}_{-}^{\dagger}(\bm{k})\mathcal{O}_{-}(\bm{k}')\nonumber \\
= & -2i(k_{x}k_{y}'-k_{x}'k_{y})(c_{\bm{k}\uparrow}^{\dagger}c_{-\bm{k}\downarrow}^{\dagger}+c_{\bm{k}\downarrow}^{\dagger}c_{-\bm{k}\uparrow}^{\dagger})(c_{-\bm{k}'\downarrow}c_{\bm{k}'\uparrow}+c_{-\bm{k}'\uparrow}c_{\bm{k}'\downarrow})\nonumber \\
= & -2i\hat{z}\cdot(\bm{k}\times\bm{k}')(c_{\bm{k}\uparrow}^{\dagger}c_{-\bm{k}\downarrow}^{\dagger}+c_{\bm{k}\downarrow}^{\dagger}c_{-\bm{k}\uparrow}^{\dagger})(c_{-\bm{k}'\downarrow}c_{\bm{k}'\uparrow}+c_{-\bm{k}'\uparrow}c_{\bm{k}'\downarrow}). 
\end{align}
It obviously breaks time-reversal symmetry, and if the Hamiltonian has a term proportional to $i\hat{z}\cdot(\bm{k}\times\bm{k}')$, it lifts the degeneracy between states with $l=\pm 1$. 
Indeed, the off-diagonal terms in Eq.~\eqref{eq:matrix} have this form. This is because the Chern-Simons term makes the density-current correlation $\langle a_0 a_1 \rangle$ finite and hence the propagators $D_{\pm,01}$ and $D_{\pm,10}$. Also we note that the off-diagonal components of Eq.~\eqref{eq:propagator} have odd powers of $\tilde{\phi}$, which indicates the violation of time-reversal symmetry. 

A comment on the imaginary effective interaction might be useful. Actually it guarantees the Hermiticity of the Hamiltonian. 
If we consider the Hermitian conjugate of Eq.~\eqref{eq:eff_int}, we obtain 
\begin{align}
&\int_{k,k',q} \mathcal{V}^\dagger \notag \\
=&\frac{1}{2}\sum_{s_1 s_2 s_3 s_4} \int_{k,k',q} \left[ V_{s_1 s_2 s_3 s_4}^{\text{eff}}(k,k',q) \right]^*
\psi_{s_4}^{\dagger}(k) \psi_{s_3}^\dagger (k') \psi_{s_2}(k'-q) \psi_{s_1} (k+q) \notag \\
=& \frac{1}{2}\sum_{s_3 s_4 s_1 s_2} \int_{k,k',q}  \left[ V_{s_3 s_4 s_1 s_2}^{\text{eff}}(k'-q,k+q,q) \right]^*
\psi_{s_1}^{\dagger}(k+q) \psi_{s_2}^\dagger (k'-q) \psi_{s_3}(k') \psi_{s_4} (k), 
\end{align}
where we define $k=(\bm{k},i\epsilon_n)$, $k'=(\bm{k}',i\epsilon_n')$, $q=(\bm{q},i\omega_m)$, and $\int_k =T\sum_{\epsilon_n} \int\frac{d^2k}{(2\pi)^2}$ etc.\ to simplify the notation. 
Therefore, the following equality holds for the effective Hamiltonian
to be hermite;
\begin{equation}
V_{s_1 s_2 s_3 s_4}^{\text{eff}}(k,k',q)=\left[V_{s_3 s_4 s_1 s_2}^{\text{eff}}(k'-q,k+q,q)\right]^{*}.
\end{equation}

\section{BCS theory}

It is indicative to mention an application of the BCS theory to the present model. 
We consider the Hamiltonian 
\begin{align}
H & =\sum_{\bm{k}s}\xi_{\bm{k}}c_{\bm{k}s}^{\dagger}c_{\bm{k}s}+\frac{1}{2}\sum_{\bm{k}\bm{k}'}\sum_{s_{1}s_{2}s_{3}s_{4}}V_{\bm{k}\bm{k}',s_{1}s_{2}s_{3}s_{4}}c_{\bm{k}s_{1}}^{\dagger}c_{-\bm{k}s_{2}}^{\dagger}c_{-\bm{k}'s_{3}}c_{\bm{k}'s_{4}},
\end{align}
where $\xi_{\bm{k}}=\epsilon_{\bm{k}}-\mu$
and $V_{\bm{k}\bm{k'},s_{1}s_{2}s_{3}s_{4}}$ is the interaction in
the Cooper channel. (The notation here is slightly different from
the other sections. $\bm{k}'$ is a wave vector of an out-going particle,
which is $\bm{k}+\bm{q}$ in the other sections.) Note frequency dependence is neglected in the BCS theory. We define the gap function
$\Delta_{\bm{k},ss'}$ as
\begin{equation}
\Delta_{\bm{k},ss'}=-\sum_{\bm{k}}\sum_{s_{1}s_{2}}V_{\bm{k}\bm{k}',ss's_{2}s_{1}}\langle c_{\bm{k}'s_{1}}c_{-\bm{k}'s_{2}}\rangle.
\end{equation}
In general, the gap function is written as 
\begin{equation}
\hat{\Delta}_{\bm{k}}=\left(\Delta_{\bm{k}}\right)_{ss'}=\Delta\left(\varphi(\bm{k})+\vec{d}(\bm{k})\cdot\vec{\sigma}\right)(i\sigma_{y})
\end{equation}
with $\varphi(\bm{k})=\varphi(-\bm{k})$ (spin-singlet) and $\vec{d}(\bm{k})=-\vec{d}(-\bm{k})$
(spin-triplet). For a unitary state, the gap equation is 
\begin{equation}
\Delta_{\bm{k},s_{1}s_{2}}=-\sum_{\bm{k}'}\sum_{s_{3}s_{4}}V_{\bm{k}\bm{k}',s_{1}s_{2}s_{3}s_{4}}\frac{\Delta_{\bm{k}',s_{4}s_{3}}}{2E_{\bm{k}'}}\tanh\left(\frac{E_{\bm{k}'}}{2T}\right),
\end{equation}
with $E_{\bm{k}}=\sqrt{\xi_{\bm{k}}+|\Delta_{\bm{k}}|^{2}}$ and $|\Delta_{\bm{k}}|^{2}=\frac{1}{2}\text{tr}(\hat{\Delta}_{\bm{k}}^{\dagger}\hat{\Delta}_{\bm{k}})$.
The spin-dependent interaction $V_{\bm{k}\bm{k}',s_{1}s_{2}s_{3}s_{4}}$
can be decomposed as 
\begin{equation}
V_{\bm{k}\bm{k}',s_{1}s_{2}s_{3}s_{4}}=J_{\bm{k}\bm{k}'}^{0}(\sigma_{0})_{s_{1}s_{4}}(\sigma_{0})_{s_{2}s_{3}}+\sum_{\alpha=x,y,z}J_{\bm{k}\bm{k}'}^{\alpha}(\sigma_{\alpha})_{s_{1}s_{4}}(\sigma_{\alpha})_{s_{2}s_{3}}.
\end{equation}
Using $J^{0}$ and $J^{\alpha}$, the gap equation become
\begin{gather}
\varphi(\bm{k})=-\sum_{\bm{k}'}\Bigl(J_{\bm{k}\bm{k}'}^{0}-\sum_{\alpha}J_{\bm{k}\bm{k}'}^{\alpha}\Bigr)\frac{\varphi(\bm{k}')}{2E_{\bm{k}'}}\tanh\left(\frac{E_{\bm{k}'}}{2T}\right),\\
d^{\alpha}(\bm{k})=-\sum_{\bm{k}'}\Bigl(J_{\bm{k}\bm{k}'}^{0}-J_{\bm{k}\bm{k}'}^{\alpha}+\sum_{\beta\neq\alpha}J_{\bm{k}\bm{k}'}^{\beta}\Bigr)\frac{d^{\alpha}(\bm{k}')}{2E_{\bm{k}'}}\tanh\left(\frac{E_{\bm{k}'}}{2T}\right).
\end{gather}
We assume that the gap function is much smaller compared to the Fermi
energy ($|\Delta_{\bm{k}}|\ll\epsilon_{F}$), and hence we can approximate
the gap function $\Delta_{\bm{k}}$ to be finite only on the Fermi
surface ($|\bm{k}|=k_{F}$). 

For the present model, the gap function is determined by 
\begin{align}
\begin{cases}
\phi(\bm{k}) = e^{il\theta_{\bm{k}}} & (l\text{: even}), \\
\vec{d}(\bm{k}) = e^{il\theta_{\bm{k}}} \hat{z} & (l\text{: odd}). 
\end{cases}
\end{align}
With the layer indices associated with spins, and even and odd $l$ states correspond to  spin-singlet and spin-triplet states. However, the layer indices as pseudospins does not have $SU(2)$ symmetry but only $U(1)$ symmetry, which corresponds to the rotation in the $xy$-plane, since the top and bottom layers have a physical meaning. The spin-singlet pairings naturally give interlayer pairings, whereas the spin-triplet states include both intralayer and interlayer pairings. We restrict our analysis to interlayer paired state, which forces $\vec{d}\parallel \hat{z}$. Note that the spin-triplet states so defined are unitary states since $\vec{d}(\bm{k})\times\vec{d}^{*}(\bm{k})=0$.

The gap equation for an $l$-wave pairing is
\begin{gather}
e^{il\theta_{\bm{k}}} =-\sum_{\bm{k}'}\left(J_{\bm{k}\bm{k}'}^{0}-J_{\bm{k}\bm{k}'}^{z}\right)\frac{e^{il\theta_{\bm{k}'}}}{2E_{\bm{k}'}}\tanh\left(\frac{E_{\bm{k}'}}{2T}\right) 
\end{gather}
for any $l$. 
If we extract divergent terms in the gauge propagator at $\omega_m = 0$, the effective interaction in the Cooper channel is given by 
\begin{gather}
J_{\bm{k}\bm{k}'}^{0} \approx -\frac{(\bm{k}\times\bm{k}')^{2}}{2m^{*2}|\bm{k}-\bm{k}'|^{2}}D_{+,11}(|\bm{k}-\bm{k}'|),\\
J_{\bm{k}\bm{k}'}^{z} \approx -\frac{(\bm{k}\times\bm{k}')^{2}}{2m^{*2}|\bm{k}-\bm{k}'|^{2}}D_{-,11}(|\bm{k}-\bm{k}'|)+i\frac{\hat{z}\cdot(\bm{k}\times\bm{k}')}{m^{*}|\bm{k}-\bm{k}'|}D_{-,01}(|\bm{k}-\bm{k}'|). 
\end{gather}
Linearizing the gap equations yields equations to determine the transition
temperature $T_{c}$. However,
the integrations over $\bm{k}'$ suffer from divergences at small
$q=|\bm{k}-\bm{k}'|$, and the gap equations in the BCS theory do
not have any appropriate cutoff to avoid the divergences. This is
an artifact of the BCS theory, which is originally based on an on-site 
attractive interaction and neglects frequency
dependence of the interaction. We will consider the Eliashberg theory in the next section. 

\section{Eliashberg theory}

To remove the divergence in the treatment with the BCS theory, in this section, we consider the Eliashberg theory to see which pairing is stable; i.e., we include the finite frequency contributions. 
The effective interaction~\eqref{eq:eff_int} appears in the action in the form 
\begin{align}
 & \frac{1}{2} \int_{k,k',q} V_{s_{1}s_{2}s_{3}s_{4}}^{\text{eff}}(k,k',q)c_{s_{1}}^{\dagger}(k+q) c_{s_{2}}^{\dagger} (k'-q) c_{s_{3}} (k') c_{s_{4}} (k) \nonumber \\
= & \frac{1}{2}\left(\frac{1}{2}\right)^{2} \int_{k,k',q} \Bigl\{
V_{s_{1}s_{2}s_{3}s_{4}}^{\text{eff}}(k,k',q) \left[c_{s_{1}}^{\dagger}(k+q) c_{s_{4}} (k) \right] \left[c_{s_{2}}^{\dagger}(k'-q) c_{s_{3}}(k') \right]\nonumber \\
& \quad -V_{s_{4}s_{2}s_{3}s_{1}}^{\text{eff}}(-k-q,k',q) \left[c_{s_{1}} (-k-q) c_{s_{4}}^{\dagger} (-k) \right] \left[c_{s_{2}}^{\dagger} (k'-q) c_{s_{3}} (k') \right]\nonumber \\
& \quad -V_{s_{1}s_{3}s_{2}s_{4}}^{\text{eff}}(k,-k'+q,q) \left[c_{s_{1}}^{\dagger} (k+q) c_{s_{4}}(k) \right] \left[ c_{s_{2}} (-k'+q) c_{s_{3}}^{\dagger} (-k')\right] \nonumber \\
& \quad +V_{s_{4}s_{3}s_{2}s_{1}}^{\text{eff}}(-k-q,-k'+q,q) \left[c_{s_{1}} (-k-q) c_{s_{4}}^{\dagger}(k) \right] \left[c_{s_{2}} (-k'+q) c_{s_{3}}^{\dagger} (-k') \right] \Bigr\}. 
\end{align}
By considering it in the Nambu space by using the four-component spinor \textbf{$\Psi(k)=[c_{\uparrow}(k), c_{\downarrow}(k), c_{\uparrow}^{\dagger}(-k), c_{\downarrow}^{\dagger}(-k)]^{T}$} and from the property
of the effective interaction $V_{s_{1}s_{2}s_{3}s_{4}}^{\text{eff}}(k,k',q)$,
the equation above can be written as 
\begin{align}
 & \frac{1}{2} \int_{k,k',q} V_{s_{1}s_{2}s_{3}s_{4}}^{\text{eff}}(k,k',q) c_{s_{1}}^{\dagger}(k+q) c_{s_{2}}^{\dagger} (k'-q) c_{s_{3}} (k') c_{s_{4}} (k) \nonumber \\
= & \frac{1}{2} \int_{k,k',q} \sum_{\mu,\nu=0,1}M_{\mu\nu}(\bm{k},\bm{k}',\hat{\bm{q}})\left[D_{+,\mu\nu}(q)(\sigma_{0})_{s_{1}s_{4}}(\sigma_{0})_{s_{2}s_{3}}+D_{-,\mu\nu}(q)(\sigma_{3})_{s_{1}s_{4}}(\sigma_{3})_{s_{2}s_{3}}\right]\nonumber \\
 & \times\left[\Psi_{k+q,s_{1}\tau_{1}}^{\dagger}\left(\frac{\tau_{(\mu)}}{2}\right)_{\tau_{1}\tau_{4}}\Psi_{k,s_{4}\tau_{4}}\right]\left[\Psi_{k'-q,s_{2}\tau_{2}}^{\dagger}\left(\frac{\tau_{(\nu)}}{2}\right)_{\tau_{2}\tau_{3}}\Psi_{k',s_{3}\tau_{3}}\right], 
\end{align}
with 
\begin{equation}
\tau_{(\mu)}=\begin{cases}
\tau_{3} & (\mu=0),\\
\tau_{0} & (\mu=1). 
\end{cases}
\end{equation}
The Pauli matrix $\tau_\alpha$ $(\alpha=0,...,3)$ acts on the Nambu space. 

Now we can write the Eliashberg equation in a simple way as 
\begin{equation}
\tilde{\Sigma}_{\rho\rho'}(k)=-T\sum_{\omega_{m}} \int\frac{d^2q}{(2\pi)^2} \tilde{G}_{\rho_{1}\rho_{2}}(k+q)\tilde{V}_{\rho_{2}\rho\rho_{1}\rho'}^{\text{eff}}(k,k+q,q), 
\end{equation}
where we define 
\begin{gather}
\tilde{G}_{\rho\rho'}(k)=-\frac{1}{Z_{n}^{2}\epsilon_{n}^{2}+\xi_{\bm{k}}^{2}+|\phi (k)|^{2}}\begin{pmatrix}
(i\epsilon_{n}Z_n+\xi_{\bm{k}})\sigma_{0} & \hat{\phi}(k)\\
\hat{\phi}^{\dagger}(k) & (i\epsilon_{n}Z_n -\xi_{\bm{k}})\sigma_{0}
\end{pmatrix}_{\tau\tau'}\\
\tilde{\Sigma}_{\rho\rho'}(k)=\begin{pmatrix}\left[1-Z_{n}\right](i\epsilon_{n})\sigma_{0} & \hat{\phi}(k)\\
\hat{\phi}^{\dagger}(k) & \left[1-Z_{n}\right](i\epsilon_{n})\sigma_{0}
\end{pmatrix}_{\tau\tau'},
\label{eq:green}\\
\tilde{V}_{\rho_{1}\rho_{2}\rho_{3}\rho_{4}}^{\text{eff}}(k,k',q)=\sum_{\mu,\nu}M_{\mu\nu}(\bm{k},\bm{k}',\hat{\bm{q}})\left[D_{+,\mu\nu}(q)(\sigma_{0}\tau_{(\mu)})_{\rho_{1}\rho_{4}}(\sigma_{0}\tau_{(\nu)})_{\rho_{2}\rho_{3}}+D_{-,\mu\nu}(q)(\sigma_{3}\tau_{(\mu)})_{\rho_{1}\rho_{4}}(\sigma_{3}\tau_{(\nu)})_{\rho_{2}\rho_{3}}\right],
\end{gather}
with $|\phi (k)|^{2}=\frac{1}{2}\text{tr}[ \hat{\phi}^{\dagger}(k) \hat{\phi} (k) ]$
and $\rho=(s,\tau)$. The gap function $\hat{\Delta} (k)$ is given
by $\hat{\Delta} (k)=\hat{\phi} (k) /Z_{n}$. 
We consider interlayer paired states, i.e., the anomalous self-energy $\hat{\phi} (k) $ should have  the form
\begin{equation}
\hat{\phi} (k)  = 
\begin{cases}
\phi_n^{(l)} (i\sigma_2) e^{il\theta_{\bm{k}}} & (l\text{: even}), \\
\phi_n^{(l)} (i\sigma_3 \sigma_2) e^{il\theta_{\bm{k}}} & (l\text{: odd}).
\end{cases}
\end{equation}
Then the Eliashberg equations for $\phi_n^{(l)}$ and $Z_{k}$ become
\begin{gather}
(1-Z_{n})\epsilon_{n}= -T\sum_{\omega_{m}} \int\frac{d^2q}{(2\pi)^2} \frac{Z_{n+m}(\epsilon_{n}+\omega_{m})}{Z_{n+m}^{2}(\epsilon_{n}+\omega_{m})^{2}+\xi_{\bm{k}+\bm{q}}^{2}+|\phi_{n+m}^{(l)}|^{2}} V_{\text{ex}}(\bm{k}, \bm{q},i\omega_{m}),\\
\phi_n^{(l)} e^{il\theta_{\bm{k}}}= -T\sum_{\omega_{m}} \int\frac{d^2q}{(2\pi)^2} \frac{\phi_{n+m}^{(l)} e^{il\theta_{\bm{k}+\bm{q}}}}{Z_{n+m}^{2}(\epsilon_{n}+\omega_{m})^{2}+\xi_{\bm{k}+\bm{q}}^{2}+|\phi_{n+m}^{(l)}|^{2}} V_{c}(\bm{k}, \bm{q},i\omega_{m}),
\label{eq:el2}
\end{gather}
with 
\begin{align}
V_{\text{ex}}(\bm{k}, \bm{q},i\omega_{m}) & = -\sum_{\alpha=\pm}\sum_{\mu\nu}M_{\mu\nu}(\bm{k},\bm{k}+\bm{q},\hat{\bm{q}})D_{\alpha,\mu\nu}(q,i\omega_{m})\nonumber \\
 & = -\frac{1}{2}\left[D_{+,00}(q,i\omega_{m})+D_{-,00}(q,i\omega_{m})\right] - \frac{(\hat{\bm{q}}\times\bm{k})^{2}}{2m^{*2}}\left[D_{+,11}(q,i\omega_{m})+D_{-,11}(q,i\omega_{m})\right],
\end{align}
\begin{align}
V_{c}(\bm{k}, \bm{q},i\omega_{m}) & =\sum_{\alpha=\pm}\sum_{\mu\nu}M_{\mu\nu}(\bm{k},-\bm{k}-\bm{q},\hat{\bm{q}}) (-1)^{\alpha}D_{\alpha,\mu\nu}(q,i\omega_{m})\nonumber \\
 & = \frac{1}{2}\left[D_{+,00}(q,i\omega_{m})-D_{-,00}(q,i\omega_{m})\right] +i\frac{\hat{z}\cdot(\hat{\bm{q}}\times\bm{k})}{m^{*}}\left[D_{+,01}(q,i\omega_{m})-D_{-,01}(q,i\omega_{m})\right]\nonumber \\
 & \quad -\frac{(\hat{\bm{q}}\times\bm{k})^{2}}{2m^{*2}}\left[D_{+,11}(q,i\omega_{m})-D_{-,11}(q,i\omega_{m})\right]. 
\end{align}
$(-1)^\alpha$ means $\pm1$ for $\alpha=\pm$.  
Note that $V_\text{ex}$ corresponds to the exchange interaction and $V_c$ to the interaction in the Cooper channel.

We assume that the gap function is much smaller than the Fermi
energy, i.e., $Z_{n+m}^{2}(\epsilon_{n}+\omega_{m})^{2}+|\phi_{n+m}|^{2}\ll\xi_{\bm{k}+\bm{q}}^{2}$. Then we can put $|\bm{k}|=k_F$, and the Eliashberg equations become 
\begin{gather}
\left(1-Z_{n}\right)\epsilon_{n}= -\pi T\sum_{\omega_{m}}\frac{Z_{n+m}(\epsilon_{n}+\omega_{m})}{\sqrt{Z_{n+m}^{2}(\epsilon_{n}+\omega_{m})^{2}+|\phi_{n+m}^{(l)}|^{2}}}\int \frac{d^2q}{(2\pi)^2}\delta\left(\xi_{\bm{k}+\bm{q}}\right)V_{\text{ex}}(\bm{k},\bm{q},i\omega_m),\\
\phi_n^{(l)} = -\pi T\sum_{\omega_{m}}\frac{\phi_{n+m}^{(l)}}{\sqrt{Z_{n+m}^{2}(\epsilon_{n}+\omega_{m})^{2}+|\phi_{n+m}^{(l)}|^{2}}}\int \frac{d^2q}{(2\pi)^2}\delta\left(\xi_{\bm{k}+\bm{q}}\right) V_{c}(\bm{k},\bm{q},i\omega_m)  \left( 1+ \frac{q}{k_F} e^{i(\theta_{\bm{q}}-\theta_{\bm{k}})} \right)^l .
\end{gather}
Now we define the effective coupling constants $\lambda_{Z,m}$ and $\lambda_{\phi,m}^{(l)}$ by 
\begin{gather}
\lambda_{Z,m} = \int \frac{d^2q}{(2\pi)^2} \delta(\xi_{\bm{k}+\bm{q}}) V_\text{ex} (\bm{k},\bm{q},i\omega_m),  
\label{eq:eff_Z} \\
\lambda_{\phi,m}^{(l)} = \int\frac{d^2q}{(2\pi)^2} \delta(\xi_{\bm{k}+\bm{q}}) V_c (\bm{k},\bm{q},i\omega_m) \left( 1+ \frac{q}{k_F} e^{i(\theta_{\bm{q}}-\theta_{\bm{k}})} \right)^l, 
\label{eq:eff_phi}
\end{gather}
which makes the Eliashberg equations 
\begin{gather}
\left(1-Z_{n}\right)\epsilon_{n}= -\pi T\sum_{\omega_{m}}\frac{\lambda_{Z,m} Z_{n+m}(\epsilon_{n}+\omega_{m})}{\sqrt{Z_{n+m}^{2}(\epsilon_{n}+\omega_{m})^{2}+|\phi_{n+m}^{(l)}|^{2}}},\\
\phi_n^{(l)} = -\pi T\sum_{\omega_{m}}\frac{\lambda_{\phi,m}^{(l)} \phi_{n+m}^{(l)}}{\sqrt{Z_{n+m}^{2}(\epsilon_{n}+\omega_{m})^{2}+|\phi_{n+m}^{(l)}|^{2}}}.
\end{gather}
The angular integrations in Eqs.~\eqref{eq:eff_Z} and \eqref{eq:eff_phi} can be performed analytically, to become 
\begin{align}
\label{eq:lambda_Z}
\lambda_{Z,m} = & \frac{1}{(2\pi)^2} \frac{m^*}{k_F} \int_0^{2k_F} dq \Biggl\{ 
-\frac{1}{\sqrt{1-\left(\dfrac{q}{2k_F}\right)^2}} [D_{+,00}(q,i\omega_m)+D_{-,00}(q,i\omega_m)] \notag \\
& \hspace{80pt} - \frac{k_F^2}{m^{*2}} \sqrt{1-\left(\frac{q}{2k_F}\right)^2} [D_{+,11}(q,i\omega_m)+D_{-,11}(q,i\omega_m)] \Biggr\},
\end{align}
\begin{align}
\label{eq:lambda_phi}
\lambda_{\phi,m}^{(l)} = & \frac{1}{(2\pi)^2} \frac{m^*}{k_F} \int_0^{2k_F} dq \Biggl\{ 
\frac{1}{\sqrt{1-\left(\dfrac{q}{2k_F}\right)^2}} \cos\left(2l\sin^{-1}\frac{q}{2k_F}\right) [D_{+,00}(q,i\omega_m)-D_{-,00}(q,i\omega_m)] \notag \\
& \hspace{75pt} + \frac{2k_F}{m^*} \sin\left(2l\sin^{-1}\frac{q}{2k_F}\right) [D_{+,01}(q,i\omega_m)-D_{-,01}(q,i\omega_m)] \notag \\
& \hspace{75pt} - \frac{k_F^2}{m^{*2}} \sqrt{1-\left(\frac{q}{2k_F}\right)^2} \cos\left(2l\sin^{-1}\frac{q}{2k_F}\right) [D_{+,11}(q,i\omega_m)-D_{-,11}(q,i\omega_m)] \Biggr\}.
\end{align}
We note that we can obtain a single equation for the frequency-dependent part of the gap function $\Delta_n^{(l)} = \phi_n^{(l)}/Z_n$ as 
\begin{align}
\Delta_n^{(l)} & =\phi_n^{(l)} + \Delta_n^{(l)} (1-Z_{n}) \nonumber \\
 & = -\pi T\sum_{\omega_{m}}\frac{1}{\sqrt{(\epsilon_{n}+\omega_{m})^{2}+|\Delta_{n+m}^{(l)}|^{2}}} \left( \lambda_{\phi,m}^{(l)} \Delta_{n+m}^{(l)} + \lambda_{Z,m} \Delta_n^{(l)} \frac{\epsilon_n+\omega_m}{\epsilon_n} \right).
\end{align}

We define dimensionless quantities as follows: 
\begin{gather}
r_{c}=\frac{\text{Coulomb energy}}{\text{kinetic energy}}=\frac{e^{2}/\varepsilon l_{0}}{\epsilon_{F}}=\frac{e^{2}k_{F}}{\varepsilon\epsilon_{F}}\sqrt{\frac{|\phi|}{2}},\\
r_{d}=k_{F}d.
\end{gather}
As observed Fig.~2 in the main text, $\lambda_{\phi,m}^{(+1)}$ has the largest negative value at any frequency for $\nu=\frac{1}{2}+\frac{1}{2}$ and $\frac{1}{4}+\frac{1}{4}$. For $\nu=\frac{1}{6}+\frac{1}{6}$, $\lambda_{\phi,m}^{(0)}$ is smallest at low frequencies. 
To make the $q$-integrations in Eq.~\eqref{eq:lambda_phi} finite, we need to introduce a cutoff momentum $q_c$, which we will explain later.

\subsection{Small momentum expansion}

Now we consider the expansions of the effective coupling constants $\lambda_{Z,m}$ and $\lambda_{\phi,m}^{(l)}$ for $|\omega_m|/\epsilon_{F}\ll (q/k_{F})^2 \ll1$ and $q\ll d^{-1}$. They explain the behavior of the effective coupling constants for small frequencies. 
We take up to $q^2$ terms in the numerators and denominators in the gauge propagator $D_{\pm,\mu\nu}$:
\begin{gather}
D_{+,\mu\nu}(q,i\omega_m)  \approx
\frac{1}{\chi_+ q + \chi'_+ q^2}
\begin{pmatrix}
\frac{2\pi}{m^*} \left(\frac{e^2}{\pi\varepsilon\tilde{\phi}^{2}}q + \chi_d q^2 \right) & \frac{q}{m^* \tilde{\phi}}\\
\frac{q}{m^* \tilde{\phi}} & -1
\end{pmatrix}, \\
D_{-,\mu\nu}(q,i\omega_m)  \approx
\frac{1}{\chi_- q^{2}}
\begin{pmatrix}
\frac{2\pi}{m^{*}} \left(\chi_d q^2 +\frac{e^2 d}{2\pi\varepsilon\tilde{\phi}^2} q^2 \right) & \frac{q}{m^* \tilde{\phi}}\\
\frac{q}{m^* \tilde{\phi}} & -1
\end{pmatrix},
\end{gather}
which are to be compared with Eqs.~\eqref{eq:dp} and \eqref{eq:dm}. $\chi_+$, $\chi'_+$, and $\chi_-$ are defined by 
\begin{gather}
\chi_+ = \frac{e^2}{\pi\varepsilon\tilde{\phi}} + \frac{k_F}{2\pi} \frac{|\omega_m|}{q^2}, \quad
\chi'_+ = \chi_d + \frac{1}{2\pi m^* \tilde{\phi}^2}, \quad
\chi_- = \tilde{\chi}_d + \frac{k_F}{2\pi} \frac{|\omega_m|}{q^3}, 
\end{gather}
where $|\omega_m|/q^2$ or $|\omega_m|/q^3$ work as cutoffs for small $q$. 
Then Eqs~\eqref{eq:lambda_Z} and \eqref{eq:lambda_phi} become
\begin{gather}
\lambda_{Z,m} = 
\frac{1}{(2\pi)^2} \frac{k_F}{m^*} \int_0^{2k_F} dq \left\{ \frac{1}{\chi_- q^2} + \frac{1}{\chi_+ q} + \left[ -\frac{5}{24 \chi_- k_F^2} -\frac{e^2 m^* d}{\chi_- k_F^2 \varepsilon \tilde{\phi}^2} -\frac{2e^2 m^*}{\chi_+ k_F^2 \varepsilon \tilde{\phi}^2} -\frac{\chi'_+}{\chi_+^2} \right] + O(q) \right\}, \\
\lambda_{\phi,m}^{(l)} =
\frac{1}{(2\pi)^2} \frac{k_F}{m^*} \int_0^{2k_F} dq \left\{ - \frac{1}{\chi_- q^2} + \frac{1}{\chi_+ q} + \left[ \frac{13}{24 \chi_- k_F^2} -\frac{e^2 m^* d}{\chi_- k_F^2 \varepsilon \tilde{\phi}^2} +\frac{1}{2\chi_- k_F^2}\left( l^2 -\frac{4l}{\tilde{\phi}} \right) +\frac{2e^2 m^*}{\chi_+ k_F^2 \varepsilon \tilde{\phi}^2} -\frac{\chi'_+}{\chi_+^2} \right] + O(q) \right\}.
\end{gather}
The first two terms in the expansions are divergent, but they have cutoffs with finite frequency $|\omega_m|$. The pairing symmetry dependent part is found at $q^0$ order, which is calculated safely without any singularity. 


\begin{figure}
\includegraphics[width=0.65\hsize]{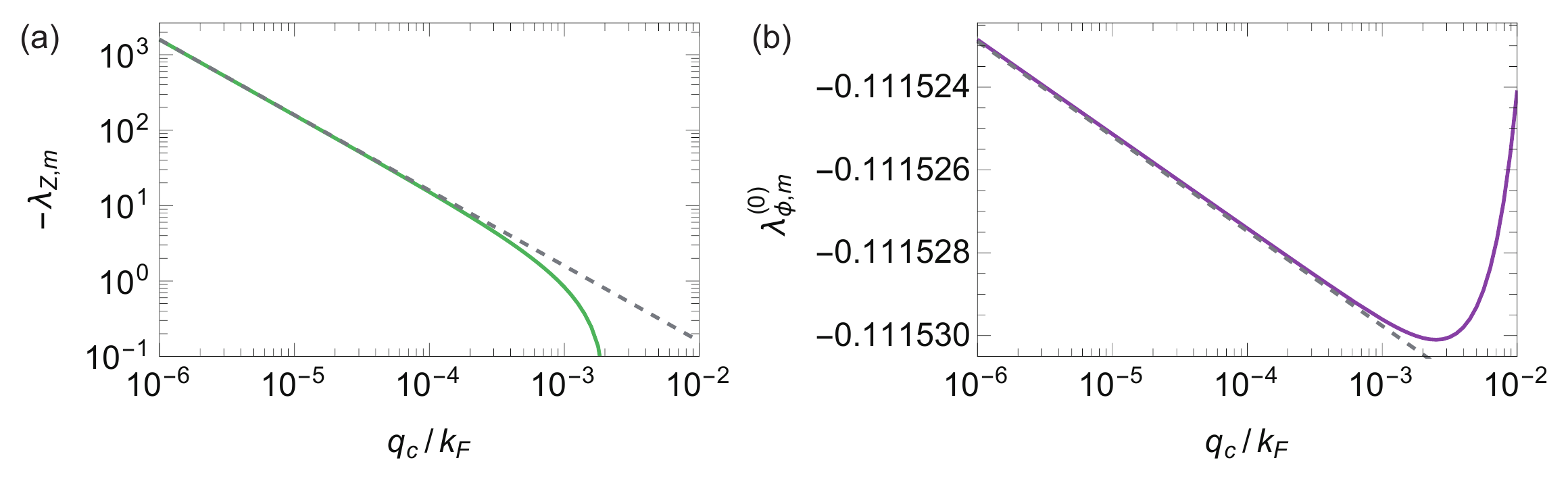}
\caption{
Asymptotic behavior of the effective coupling constants (a) $\lambda_{Z,m}$ and (b) $\lambda_{\phi,m}^{(0)}$ at filling $\nu=\frac{1}{2}+\frac{1}{2}$. 
We set the interaction strength $r_{c}=1$, layer spacing $r_{d}=1$, and cutoff $q_{c}/k_{F}=10^{-5}$.
The dashed lines represent the asymptotic form Eq.~\eqref{eq:asymptotic}.
}
\label{fig:l_freq}
\end{figure}

Asymptotic forms for small $|\omega_m|$ are calculated by using the first terms of the expansions, and the effective coupling constants become
\begin{align}
\lambda_{Z,m} \approx -\lambda_{\phi,m}^{(l)} \approx \frac{1}{(2\pi)^2} \frac{k_F}{m^*} \int dq \frac{1}{\tilde{\chi}_d q^2 +\frac{k_F}{2\pi} \frac{|\omega_m|}{q}}, 
\end{align}
where the integration focuses on the small $q$ region. It requires a lower cutoff $q_L$, and for finite $\omega_m$ it is given by 
\begin{equation}
q_L \approx \left( \frac{k_F}{2\pi\tilde{\chi}_d} |\omega_m| \right)^{1/3}. 
\end{equation}
With this $q_L$, the asymptotic form of the effective coupling constants is 
\begin{equation}
\label{eq:asymptotic}
\lambda_{Z,m} \approx -\lambda_{\phi,m}^{(l)} \approx \frac{1}{(2\pi)^2} \frac{1}{m^* \tilde{\chi}_d} \left( \frac{2\pi\tilde{\chi}_d k_F^2}{|\omega_m|} \right)^{1/3}. 
\end{equation}
Numerical results are shown in Fig.~\ref{fig:l_freq}.

\subsection{Layer spacing and effective mass dependences}

\begin{figure}
\includegraphics[width=0.7\hsize]{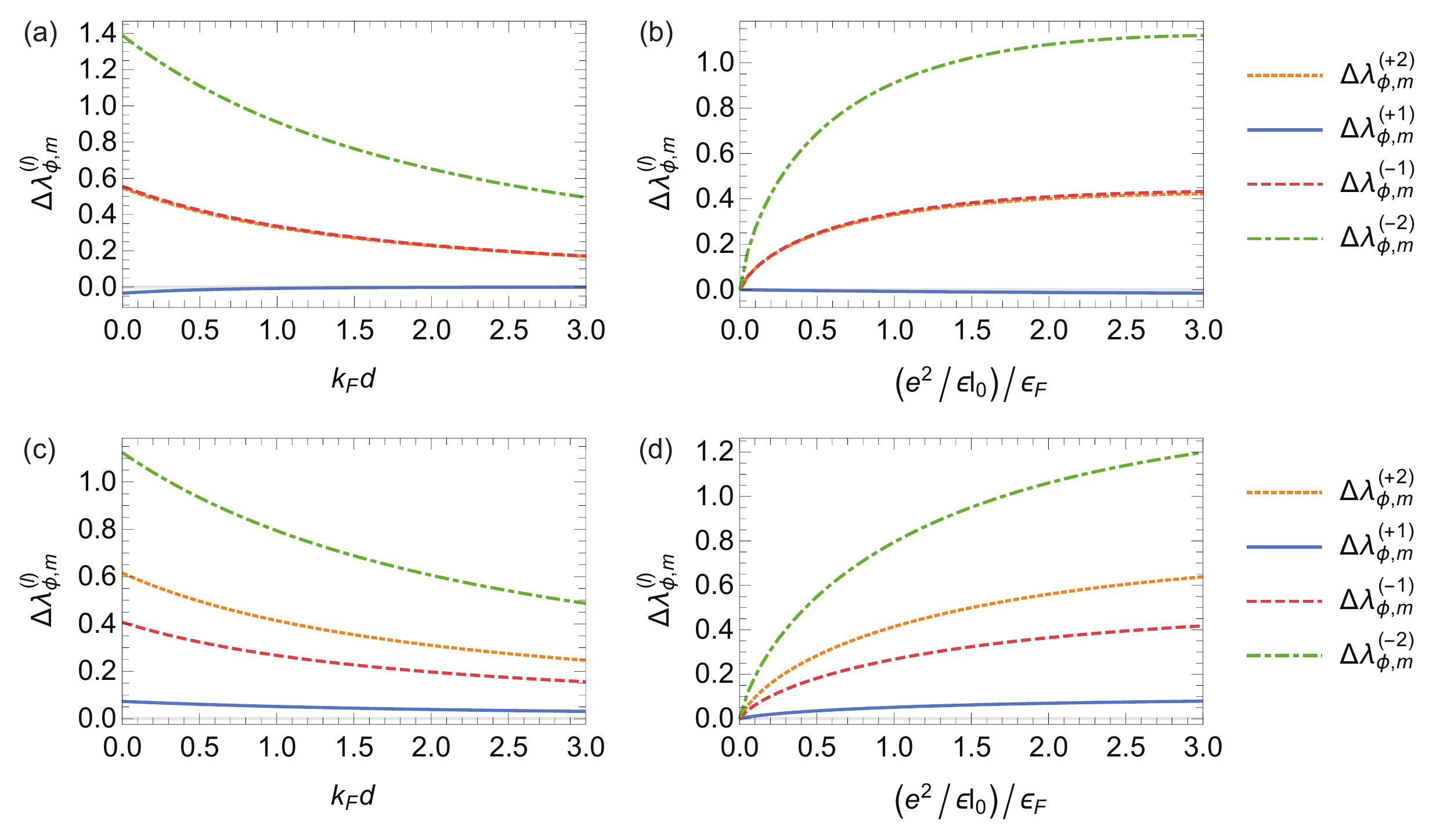}
\caption{%
(a), (c) Layer spacing $k_F d$ and (b), (d) effective mass $m^* \propto (e^2/\varepsilon l_0)/\epsilon_F$ dependences of the effective coupling constants $\Delta\lambda_{\phi,m}^{(l)}$. (a), (b) correspond to filling $\nu=\frac{1}{4}+\frac{1}{4}$, and (c), (d) correspond to $\nu=\frac{1}{6}+\frac{1}{6}$, all at $\omega_m = 0$. 
In (a) and (c), we set $r_c =1$, and $r_d =1$ in (b) and (d). 
}
\label{fig:l_fraction}
\end{figure}

The layer spacing and effective mass dependences of the effective coupling constants $\Delta\lambda_{\phi,m}^{(l)}$ for $\nu=\frac{1}{2}+\frac{1}{2}$ are shown Fig.~3 in the main text. 
Here we give the results for $\nu=\frac{1}{4}+\frac{1}{4}$ and $\frac{1}{6}+\frac{1}{6}$ (Fig.~\ref{fig:l_fraction}). The results are similar to the cases for $\nu=\frac{1}{2}+\frac{1}{2}$; the ordering of $\Delta\lambda_{\phi,m}^{(l)}$ is not changed by $k_F d$ and $(e^2/\varepsilon l_0)/\epsilon_F$ and decreasing $d$ increases $\Delta\lambda_{\phi,m}^{(l)}$. 

\subsection{Cutoff for integrations}
\label{sec:cutoff}

\begin{figure}
\includegraphics[width=0.65\hsize]{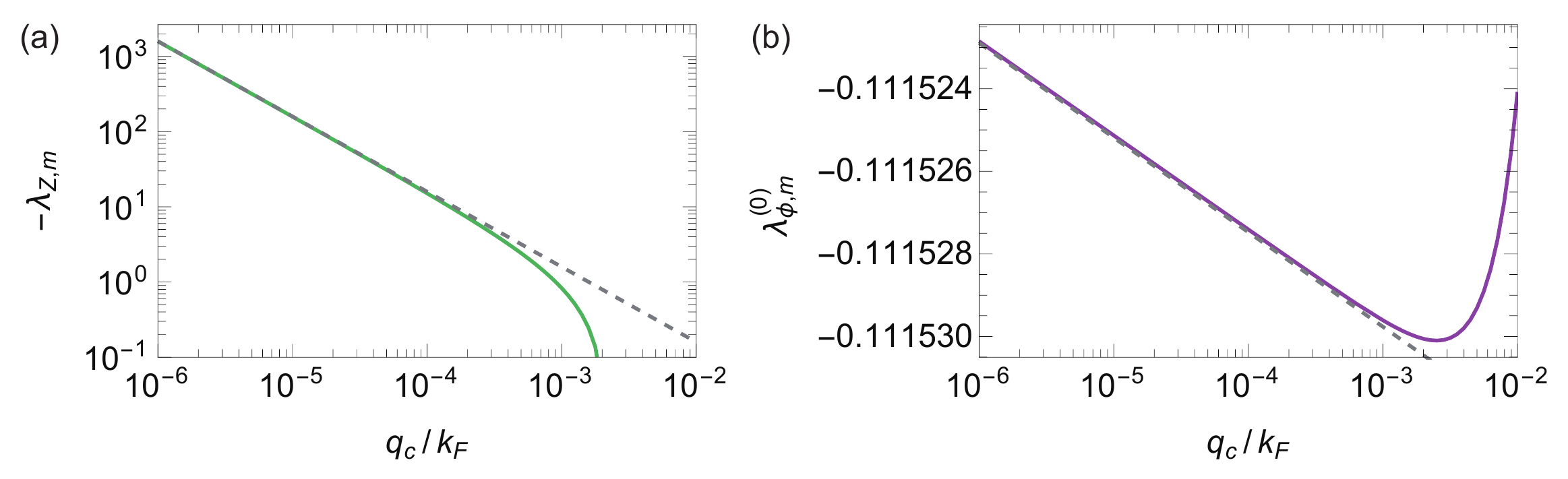}
\caption{%
Cutoff dependence of (a) $\lambda_{Z,m}$ and (b) $\lambda_{\phi,m}^{(l)}$. We set $\omega_m/\epsilon_F = 0.1$, $r_c=1$, and $r_d=1$. The dashed lines of the left and right panels correspond to the approximate forms Eqs.~\eqref{eq:lambda_Z_cut} and \eqref{eq:lambda_phi_cut}, respectively. We added a constant $-0.111536$ to the approximate form for $\lambda_{\phi,m}^{(l)}$ to fit the numerical result. 
}
\label{fig:l_cutoff}
\end{figure}

When we consider the momentum integration in Eqs.~\eqref{eq:lambda_Z} and \eqref{eq:lambda_phi} with finite frequency $|\omega_m|$, we need to use the expansion for $|\omega_m|/\epsilon_{F}\gg (q/k_{F})^2$ at smallest $q$ region. There are also singularities in the integrands appearing in this limit from the density-density components $D_{\pm,00}(q,i\omega_m)$; see Eqs.~\eqref{eq:dp_h} and \eqref{eq:dm_h}.  We need to introduce a cutoff momentum $q_c$ to avoid divergences, and then $\lambda_{Z,m}$ and $\lambda_{\phi,m}^{(l)}$ are evaluated as 
\begin{gather}
\label{eq:lambda_Z_cut}
\lambda_{Z,m} \approx -\frac{1}{(2\pi)^2} \frac{m^*}{k_F} \int_{q_c} dq [D_{+,00}(q,i\omega_m)+D_{-,00}(q,i\omega_m)] 
\approx -\frac{1}{2\pi} \frac{\omega_m^2}{\epsilon_F^2} \frac{k_F}{q_c}, \\
\label{eq:lambda_phi_cut}
\lambda_{\phi,m}^{(l)} \approx \frac{1}{(2\pi)^2} \frac{m^*}{k_F} \int_{q_c} dq [D_{+,00}(q,i\omega_m)-D_{-,00}(q,i\omega_m)] 
\approx \frac{e^2 k_F \omega_m^4}{2\pi\varepsilon\epsilon_F^5 \tilde{\phi}^5} (-\log q_c). 
\end{gather}
The cutoff dependence of $\lambda_{Z,m}$ and $\lambda_{\phi,m}^{(l)}$ is shown in Fig.~\ref{fig:l_cutoff}. 

\section{Wave functions of paired states}
\subsection{Composite fermions}

From the Green's function Eq.~\eqref{eq:green}, we can construct the effective action
\begin{equation}
S^{\text{eff}}=-\frac{1}{2}\int_{k}\Psi^{\dagger}(k)\tilde{G}^{-1}(k)\Psi(k),
\end{equation}
which is reduce to be 
\begin{equation}
S^{\text{eff}}=\int_{k}\left[\sum_{s=\uparrow,\downarrow}(-i\epsilon_{n}Z_{n}+\xi_{\bm{k}})c_{s}^{\dagger}(k)c_{s}(k)-\phi(k)c_{\uparrow}^{\dagger}(k)c_{\downarrow}^{\dagger}(-k)-\phi^{*}(k)c_{\downarrow}(-k)c_{\uparrow}(k)\right], 
\end{equation}
with $\phi(k)=\phi_{n}^{(l)}e^{il\theta_{\bm{k}}}$. 
This effective action is associated with the BCS mean-field Hamiltonian
\begin{equation}
H^{\text{mf}}=\sum_{\bm{k}}\left[\sum_{s=\uparrow,\downarrow}\xi_{\bm{k}}c_{s}^{\dagger}(k)c_{s}(k)-\Delta_{\bm{k}}c_{\uparrow}^{\dagger}(k)c_{\downarrow}^{\dagger}(-k)-\Delta_{\bm{k}}^{*}c_{\downarrow}(-k)c_{\uparrow}(k)\right]. 
\end{equation}
For the mean-field Hamiltonian, we consider the Bogoliubov transformation
\begin{gather}
\alpha_{\bm{k}\uparrow}=u_{\bm{k}}c_{\bm{k}\uparrow}-v_{\bm{k}}c_{-\bm{k}\downarrow}^{\dagger},\\
\alpha_{\bm{k}\downarrow}=u_{\bm{k}}c_{\bm{k}\downarrow}+v_{\bm{k}}c_{-\bm{k}\uparrow}^{\dagger},
\end{gather}
where $\alpha_{\bm{k}s}$ satisfies the anticommutation relations $\{\alpha_{\bm{k}s},\alpha_{\bm{k}'s'}^{\dagger}\}=\delta_{\bm{k}\bm{k}'}\delta_{ss'}$
and $\{\alpha_{\bm{k}s},\alpha_{\bm{k}'s'}\}=\{\alpha_{\bm{k}s}^{\dagger},\alpha_{\bm{k}'s'}^{\dagger}\}=0$
with the condition $|u_{\bm{k}}|^{2}+|v_{\bm{k}}|^{2}=1$. Two parameters
$u_{\bm{k}}$ and $v_{\bm{k}}$ are determined by imposing the following commutation relations 
\begin{equation}
[\alpha_{\bm{k}\uparrow},H^{\text{mf}}]=E_{\bm{k}}\alpha_{\bm{k}\uparrow},\quad[\alpha_{\bm{k}\downarrow},H^{\text{mf}}]=E_{\bm{k}}\alpha_{\bm{k}\downarrow},
\end{equation}
which yield two coupled equations
\begin{subequations}
\begin{gather}
\begin{cases}
E_{\bm{k}}u_{\bm{k}}=\xi_{\bm{k}}u_{\bm{k}}+\Delta_{\bm{k}}^{*}v_{\bm{k}}\\
E_{\bm{k}}v_{\bm{k}}=\Delta_{\bm{k}}u_{\bm{k}}-\xi_{\bm{k}}v_{\bm{k}}
\end{cases}
\end{gather}
and 
\begin{equation}
\begin{cases}
E_{\bm{k}}u_{\bm{k}}=\xi_{\bm{k}}u_{\bm{k}}+\Delta_{-\bm{k}}^{*}v_{\bm{k}}\\
E_{\bm{k}}v_{\bm{k}}=\Delta_{-\bm{k}}u_{\bm{k}}-\xi_{\bm{k}}v_{\bm{k}}
\end{cases}
\end{equation}
\end{subequations}
Those equations are satisfied at the same time if the parities of $v_{\bm{k}}$ and $\Delta_{\bm{k}}$ match, and we obtain
\begin{gather}
E_{\bm{k}}=\sqrt{\xi_{\bm{k}}^{2}+|\Delta_{\bm{k}}|^{2}},\\
u_{\bm{k}}=\frac{\xi_{\bm{k}}+E_{\bm{k}}}{\sqrt{2E_{\bm{k}}(\xi_{\bm{k}}+E_{\bm{k}})}},\\
v_{\bm{k}}=\frac{\Delta_{\bm{k}}}{\sqrt{2E_{\bm{k}}(\xi_{\bm{k}}+E_{\bm{k}})}}.
\end{gather}
The Bogoliubov transformation makes the Hamiltonian diagonalized to be 
\begin{equation}
H^{\text{mf}}=\sum_{\bm{k}s}E_{\bm{k}}\alpha_{\bm{k}s}^{\dagger}\alpha_{\bm{k}s}+\text{const.}
\end{equation}
The ground state for the Hamiltonian $H^\text{mf}$ is given by 
\begin{equation}
|\Psi\rangle=\prod_{\bm{k}}\alpha_{\bm{k}\uparrow}\alpha_{-\bm{k}\downarrow}|0\rangle
\end{equation}
with $|0\rangle$ being the vacuum, because any $\alpha_{\bm{k}s}$
annihilates this state; $\alpha_{\bm{k}s}|\Psi\rangle=0$. It is rewritten
as 
\begin{equation}
|\Psi\rangle\propto\prod_{\bm{k}}(1+g_{\bm{k}}c_{\bm{k}\uparrow}^{\dagger}c_{-\bm{k}\downarrow}^{\dagger})|0\rangle,
\end{equation}
where $g_{\bm{k}}=v_{\bm{k}}/u_{\bm{k}}$. The projection onto a space with $N$ particles ($N$: even) 
gives the unnormalized
wave function
\begin{equation}
\Psi(\bm{r}_{i\uparrow},\bm{r}_{j\downarrow})=\det[g(\bm{r}_{i\uparrow},\bm{r}_{j\downarrow})],
\end{equation}
where $g(\bm{r}_{i\uparrow},\bm{r}_{j\downarrow})$ is the Fourier
transform of $g_{\bm{k}}$ 
\begin{equation}
g(\bm{r}_{i\uparrow},\bm{r}_{j\downarrow})=\frac{1}{L^{2}}\sum_{\bm{k}}g_{\bm{k}}e^{i\bm{k}\cdot(\bm{r}_{i\uparrow}-\bm{r}_{j\downarrow})}.
\end{equation}
$L^{2}$ is the area of the system. When the relative angular momentum of an interlayer pairing is $l$, we have 
\begin{equation}
g(\bm{r}) = (x+iy)^l f(r),
\label{eq:function_g}
\end{equation}
where $f$ is an arbitrary function of $r=\sqrt{x^2+y^2}$. The function $f(r)$ does not contribute to the relative angular momentum.

\subsection{Electrons}

The previous subsection focuses on the pairing of composite fermions.
The wave function of electrons includes flux attachment, or technically, singular gauge transformation, which forms a boson part in the wave function. 
As a result, the wave function of electrons $\Psi$ is composed of the boson part $\Psi_\text{B}$ and composite fermion part $\Psi_\text{CF}$:
\begin{equation}
\Psi = \Psi_\text{B} \Psi_\text{CF}.
\end{equation}
The boson part is represented by a bosonic Halperin state $(m,m,n)$
\begin{equation}
\Psi_\text{B} = \prod_{i<j} (z_i-z_j)^m \prod_{i'<j'} (w_{i'}-w_{j'})^m \prod_{r,s} (z_r-w_s)^n, 
\end{equation}
with even integers $m$ and $n$. The integers $m$ and $n$ determine the filling fraction of a layer as $1/(m+n)$, and hence the total filling is $\nu=2/(m+n)$. 
Here we introduce the complex representation of the two-dimensional coordinate $z_i=x_{i\uparrow}-iy_{i\uparrow}$ for the top layer and $w_j=x_{j\downarrow}-iy_{j\downarrow}$ for the bottom layer. 
This definition is required by the sign of $eB$; if we choose the convention with $eB<0$, the definition should be complex conjugate; $z_i \to x_{i\uparrow}+iy_{i\uparrow}$ and $w_j \to x_{j\downarrow}+iy_{j\downarrow}$. 

Our Lagrangian \eqref{eq:lagrangian} corresponds to the $(\tilde{\phi},\tilde{\phi},0)$ state, where the bosons are incoherent between layers, when the layer spacing is not very small. 
As the layer spacing decreases, we expect $n\neq 0$, which yields interlayer coherence of the bosons. 

The composite fermion part $\Psi_\text{CF}$ is dictated by using the function $g(\bm{r}_{i\uparrow}-\bm{r}_{j\downarrow})=g(z_i-w_j)$, i.e., 
\begin{equation}
g(z_i-w_j) = (z_i-w_j)^{-l} f(|z_i-w_j|), 
\end{equation}
with the phase-independent function $f$ is redefined from Eq.~\eqref{eq:function_g} to absorb $|z_i-w_j|^{2l}$. $(z_i-w_j)^{-l}$ represents the angular momentum $l$ with the definition $z_i=x_{i\uparrow}-iy_{i\uparrow}$ and $w_j=x_{j\downarrow}-iy_{j\downarrow}$. Now we have 
\begin{equation}
\Psi_\text{CF} = \det[ g(z_i-w_j) ].
\end{equation}
When the function $f(|z_i-w_j|)$ is of order of unity at large $|z_i-w_j|$, the phase is called ``weak-pairing''  and if $f(|z_i-w_j|)$ falls rapidly, say exponentially, at large distances, then the phase is ``strong-pairing'' \cite{s-RG}. Those two phases are different in topology.  We note that our analysis does not distinguish the two. 

When $f(|z_i-w_j|) = 1$, which corresponds to the weak-pairing phase, the Cauchy identity
\begin{equation}
\prod_{i<j}(z_{i}-z_{j})\prod_{k<l}(w_{k}-w_{l}) = \prod_{i,j}(z_{r}-w_{s}) \cdot \det \left(\frac{1}{z_i-w_j}\right), 
\end{equation}
leads to 
\begin{equation}
\Psi_\text{CF} = \prod_{i<j} (z_i-z_j)^l \prod_{i'<j'} (w_{i'}-w_{j'})^l \prod_{r,s} (z_r-w_s)^{-l} .
\end{equation}
A weak-pairing phase with the relative angular momentum $l$ can also be regarded as the $(l,l,-l)$ state. 

The wave function of electrons are given by the product of $\Psi_\text{B}$ and $\Psi_\text{CF}$, as 
\begin{equation}
\Psi(\{z\},\{w\})=\mathcal{P}_{\text{LLL}} \prod_{i<j} (z_i-z_j)^m \prod_{i'<j'} (w_{i'}-w_{j'})^m \prod_{r,s} (z_r-w_s)^n \cdot\det[g(z_{i},w_{j})]. 
\end{equation}
For $\nu=\frac{1}{2}+\frac{1}{2}$, our analysis focus on the case with $m=2$ and $n=0$, and find the $l=+1$ state is energetically favored. If it is a weak-pairing phase, its topological property is equivalent to the $(3,3,-1)$ state and the ground state degeneracy is eight on a torus. In contrast, when it is a strong-pairing phase, the ground state is four-fold degenerate \cite{s-Kim}. 
If we suppose that the boson part is the $(0,0,2)$ state with a small layer spacing and that the $l=+1$ state is still favored and weak-pairing, we would obtain the $(1,1,1)$ state as the ground state. 
For $\nu=\frac{1}{4}+\frac{1}{4}$, the $l=1$ state in a weak-pairing phase is analogous to the $(5,5,-1)$ state. Again, if we assume the boson part as the $(2,2,2)$ state with the $l=+1$ paired state of composite fermions, the resulting state becomes the $(3,3,1)$ state.


\begin{thebibliography}{99}
\bibitem{screen}
In an experiment, gate electrodes adjacent to CF layers could screen the long-range Coulomb interaction and suppress the charge density fluctuation. 
Such a screening effect in bilayer graphene experiments is discussed in Ref.~\cite{scr_exp}, for example.
\bibitem{triplet}
States with even $l$ are spin-singlet pairings, while those with odd $l$ are spin-triplet states, where ``spin'' corresponds to layer in the present model.  
The spin component of spin-singlet states is dictated by $i\sigma_2$, which corresponds to interlayer singlet pairing in bilayer systems. On the other hand, spin-triplet states still have spin degrees of freedom, but since we focus on interlayer pairings, their spin components are described by $\sigma_3(i\sigma_2)$, where a corresponding $d$ vector is $\bm{d}\propto e^{il\theta_{\bm{k}}} \hat{z}$.
\bibitem{definition}
The definitions of the complex representation of the coordinate $z_i$ and $w_j$ depend on the sign of a product $eB$. The definitions here correspond to the case with $eB>0$. In contrast, if we had $eB<0$, the choice would be $z_i=x_{i\uparrow}+iy_{i\uparrow}$ and $w_j=x_{j\downarrow}+iy_{j\downarrow}$.
\end{thebibliography}
\end{document}